\documentclass[article,oneauthor,pdftex,10pt,a4paper]{Definitions/mdpi-arxiv}

\usepackage{amssymb,amsfonts}

\Title{Young's experiment with entangled bipartite systems: The role of underlying quantum velocity fields}

\Author{{\'Angel S. Sanz} \orcidA{}}

\address[1]{Department of Optics, Faculty of Physical Sciences,
Universidad Complutense de Madrid, {Pza. Ciencias 1,} Ciudad Universitaria, 28040 Madrid, Spain; a.s.sanz@fis.ucm.es; Tel.: +34-91-394-4673}

\abstract{We consider the concept of velocity fields, taken from Bohmian mechanics, to investigate the
dynamical effects of entanglement in bipartite realizations of Young’s two-slit experiment.
In particular, by comparing the behavior exhibited by factorizable two-slit states
(cat-type state analogs in the position representation) with the dynamics exhibited by
a continuous-variable Bell-type maximally entangled state, we find that, while the velocity
fields associated with each particle in the separable scenario are well-defined and act
separately on each subspace, in the entangled case there is a strong deformation in the
total space that prevents this behavior.
Consequently, the trajectories for each subsystem are not constrained any longer to remain
confined within the corresponding subspace; rather, they exhibit seemingly wandering
behavior across the total space.
In this way, within the subspace associated with each particle (that is, when we trace over
the other subsystem), not only interference features are washed out, but also the so-called
Bohmian non-crossing rule\linebreak  (i.e., particle trajectories are allowed to get across the same
point at the same time).
}


\usepackage{xcolor}
\usepackage{soul}

\begin{document}


\section{Introduction}
\label{sec1}

The 2022 Nobel Prize in Physics was awarded to Alain Aspect, John F.~Clauser,\linebreak  and Anton
Zeilinger ``for experiments with entangled photons, establishing the violation of Bell
inequalities and pioneering quantum information science'' \cite{Nobel:2022}.
All these ideas started in the mid-1930s as a heated debate about a ``spooky action at a distance''
\cite{EPR:PhysRev:1935}, the separability of quantum systems~\cite{schrodinger:ProcCamPS:1935},
the impossibility of hidden variables in quantum mechanics~\cite{vonNeumann-bk:1932},\linebreak  and, in sum,
the problem of quantum measurement~\cite{zurek-bk,schlosshauer:RMP:2004,gilder-bk}.
However, entanglement has become, together with coherence, the cornerstone of modern quantum
information theory~\cite{nielsen-chuang-bk} and the quantum technologies that have emerged after
the so-called second\linebreak  quantum~revolution~\cite{milburn:PTRSLA:2003}.

The analysis of entanglement and its properties is often based on rather abstract (algebraic)
descriptions, where the fact that quantum systems go from somewhere to somewhere else in a
given time is typically not considered.
Young's two-slit experiment is an ideal candidate to render some light on the phenomenology
of how this happens, as it constitutes a paradigm in the investigation of properties related
to quantum coherence, a key element strongly connected to entanglement.
This experiment not only makes evident the reality and implications of the quantum phase
through the appearance of interference fringes, but also the gradual or total suppression of
the latter as soon as the system, diffracted by the slits, interacts with, and becomes entangled with,
other surrounding systems.
This coherence loss mechanism is what we call decoherence \cite{schlosshauer:PhysRep:2019}
(to be distinguished from the decrease of fringe visibility due to the presence of
incoherence factors \cite{sanz:PRA:2005}), which leads to behaviors analogous to those
observed in the ordinary (incoherent) classical world
\cite{giulini-bk,schlosshauer-bk:2007}.

Among the several pictures of quantum mechanics, the results of the Bohmian \linebreak one~\cite{bohm:PR:1952-1,bohm-hiley-bk,holland-bk} are particularly appealing to investigate
\textls[-10]{the above type of problem, as it puts the emphasis on the role of the phase in quantum
dynamics \cite{sanz:AJP:2012,sanz:FrontPhys:2019}.
Different questions} related to entanglement have already been considered from this point of
view in the literature \cite{plastino:Entropy:2018,tzemos:PRE:2020,tzemos:PRE:2021,tzemos:PhysScr:2021}, including those aimed at proving, or disproving, flaws in the\linebreak  approach
\cite{guay:JPA:2003,walborn:PRA:2002,brida:PRA:2003,golshani:JPA:2001,kwiat:LAS:2002,ghirardi:JStatPhys:2002,ghirardi:PRA:2004,matzkin:PRA:2011}.
Leaving aside interpretive issues, from the position that Bohmian trajectories are simply
(quantum) flux streamlines describing the spatial dispersion of the probability density along
time, we find that Bohmian mechanics becomes a suitable tool to study and understand the
evolution of quantum systems in an alternative way to that of considering the global information
supplied by the probability density.
This relies on a well-defined velocity field, determined from the relationship between the
quantum flux and the probability density (the so-called guidance condition in canonical
Bohmian mechanics \cite{holland-bk}).
As was shown in \cite{sanz:JPA:2008,sanz:FrontPhys:2019}, for instance, this velocity
field is very convenient in the analysis of the dynamical consequences intrinsically associated
with the concept of quantum coherence.

In this work, we focus on the relationship between entanglement and decoherence in
investigating the continuous coherence flow between two spatially separated subsystems with
the aid of both the (Bohmian) velocity field and the trajectories obtained from it.
To this end, we consider a realization of Young's two-slit experiment with two identical
particles, $X$ and $Y$, with a mass $m$, given that this paradigm has been used in the past
to test experimentally different aspects of Bohmian mechanics
\cite{brida:PRA:2003,kocsis:Science:2011,braverman:PRL:2013}.
Without any loss of generality, particle $X$ is taken as the reference system, which
describes the coherent superposition of the two diffracted beams in the experiment.
As for $Y$, it is taken as the ``environmental'' partner, which leads to the suppression
of the interference traits present in $X$ when both are entangled.
From the most elementary aspects of the theory of open quantum systems
\cite{breuer-bk:2002}, it is known that such suppression is an effect of $Y$ on the
projection of the state vector for the full system $XY$ onto the subspace for $X$.
In this paper, the objective is to determine the specific dynamical aspects
that make the entangled state different with respect to the factorizable one, and which can be
extrapolated to more general frameworks, regardless of whether the two subsystems are
spatially separated or they are interacting~\cite{sanz:ENTROPY:2022}.

Accordingly, the work has been organized as follows.
In Section~\ref{sec2}, we introduce and discuss basic theoretical aspects involved in the
description of continuous variable systems, including a brief overview on both localized
and delocalized states, described in terms of Gaussian wave packets, mimicking diffraction
by Gaussian slits \cite{feynman-bk1,feynman-bk1-2010}.
The results from numerical simulations illustrating the main dynamical features for both
factorizable and entangled states are shown and discussed in Section~\ref{sec3}.
Finally, the main conclusions from this work are summarized in Section~\ref{sec4}.


\section{Theoretical Aspects}
\label{sec2}


\subsection{Basic Formal Grounds}
\label{sec21}

To be self-contained, let us first introduce some general, basic considerations.
Thus, let us consider two coherently emitting sources (e.g., two slits,
two coupled fibers, etc.), with the transmitted waves traveling ahead faster
than along the transverse direction,\linebreak  i.e., the direction along which we expect
to detect the effects of coherence (interference) or to analyze their suppression
(decoherence).
This simplified phenomenological physical system has been considered previously with excellent results;
for instance, in the analysis of grating diffraction
\cite{sanz:AOP:2015}.
Accordingly, if we consider two identical particles, $X$ and $Y$, their descriptions
rely on their transverse coordinates; henceforth, denoted by $x$ and $y$,
respectively.
Additionally, to simplify the analysis, let us consider that $X$ and $Y$ are
both nonrelativistic, spinless particles of mass $m$.

In a first instance, following the standard approach, the analysis of the
dynamical implications of the coupling between $X$ and $Y$ requires a description
of the problem based on a density matrix approach.
If the joint state for $X$ and $Y$ is represented by the state vector $|\Psi (t)\rangle$,
then the density matrix in the position representation (i.e., the space described by the
transverse coordinates $x$ and $y$) for the joint system is
\begin{equation}
 \hat{\rho}(x,y;x',y'|t) = \rho(x,y;x',y'|t) |x,y \rangle \langle x',y'| ,
 \label{eq1}
\end{equation}
where
\begin{equation}
 \rho(x,y;x',y'|t) = \langle x,y|\Psi(t)\rangle \langle \Psi(t)|x',y'\rangle
  = \Psi^*(x',y'|t) \Psi(x,y|t) .
 \label{eq2}
\end{equation}
denotes the (density matrix) element in the position representation.
The diagonal elements provide us with the usual probability density, while the off-diagonal
elements provide an account of the preservation of the full-system ($XY$) quantum coherence.
For simplicity, but without loss of generality, we always work under
the hypothesis that the full system is pure and, hence, the (full-system) density matrix takes the usual form $\hat{\rho} = |\Psi (t)\rangle \langle \Psi (t)|$, which gives
rise to Equation~(\ref{eq1}) after considering the corresponding projection operators.
This is justified by the fact that, here, we are investigating the joint
dynamics.
However, the current analysis can readily be generalized to mixed states.

To quantify the effects of the coupling for pure bipartite systems, it is also common to
trace the density matrix over one of the subsystems and, then, to analyze the properties
displayed by the reduced density matrix associated with the remaining subsystem
\cite{sanz:ENTROPY:2022}.
Proceeding in this way, tracing over $Y$, for instance {{(the same description can
be achieved the other way around, tracing over $X$ instead of $Y$, without any detrimental effect
regarding the conceptual aspects that are here investigated)}}, leads to the reduced density
matrix for~$X$:
\begin{eqnarray}
 \hat{\tilde{\rho}}_X(x,x'|t) = {\rm Tr}_Y [ \hat{\rho}(x,y;x',y'|t) ]
  & = & \iint \left( \int \langle y'' | \rho(x,y,x',y')
  |x,y\rangle \langle x',y'| y'' \rangle dy'' \right) dy dy'
 \nonumber \\
  & = & \iint \rho(x,y;x',y'|t) \left( \int
  \langle y'' | x,y\rangle \langle x',y'| y'' \rangle dy'' \right)
  dy dy'
 \nonumber \\
  & = & \tilde{\rho}_X (x,x'|t) |x\rangle \langle x'| ,
 \label{eq3}
\end{eqnarray}
where
\begin{equation}
 \tilde{\rho}_X (x,x'|t) = \int \rho(x,y;x',y|t) dy
 \label{eq4}
\end{equation}
denotes the (density matrix) element for the subsystem $X$ referred to its own subspace\linebreak 
(described by the transverse coordinate $x$).
If the two subsystems are uncorrelated\linebreak (separable), the reduced density matrix
element reduces to the trivial result
\begin{equation}
 \tilde{\rho}_X (x,x'|t) \equiv \langle x| \Psi^X (t)\rangle \langle \Psi^X(t)|x'\rangle
  = \Psi^{X,*}(x',t) \Psi^X(x,t) = \rho_X (x,x'|t) ,
 \label{eq5}
\end{equation}
where $|\Psi^X(t)\rangle$ refers to the separate state vector for $X$.
Otherwise, the effects of $Y$ manifest in a more or less relevant suppression of the off-diagonal
terms, $(x,x')$, in \mbox{Equation~(\ref{eq4})}, depending on the nature and strength of the coupling between $X$
and $Y$.
This phenomenon is what we regard as decoherence, which leads to seemingly classical density
\mbox{distributions~\cite{giulini-bk}}, even though the full or joint state remains totally
quantum mechanical.

In order to obtain a hydrodynamical picture \cite{sanz:FrontPhys:2019} of the correlation effects,
there are two ways to proceed.
First, one could consider the reduced version of the problem, that is, the description enabled by
Equation~(\ref{eq3}), and, then, compute the associated reduced quantum trajectories, as is done in
{Ref.~\cite{sanz:EPJD:2007}:} 
\begin{equation}
 v^X(x,t) = \frac{1}{m} \frac{{\rm Re} \left[ \hat{p}_x \tilde{\rho}_X(x,x'|t) \right]}
  {{\rm Re} \left[ \tilde{\rho}_X(x,x'|t) \right]} \Bigg\arrowvert_{x'=x} ,
 \label{eq6}
\end{equation}
where $\hat{p}_x = -i\hbar \partial/\partial x$ is the usual momentum operator.
Note that the above transport equation describes the relationship between the quantum flux within
the subspace for $X$, namely,
$\tilde{j}(x,t) = {\rm Re} \left[ \hat{p}_x \tilde{\rho}_X(x,x'|t) \right]_{x'=x}$,
and its corresponding reduced density distribution, $\tilde{\rho}_X(x,t)$.
The use of Equation~(\ref{eq6}) thus serves only to analyze the dynamics constrained to the $X$-subspace,
neglecting any direct or explicit contribution from the subsystem $Y$,\linebreak  since such a guidance equation
only takes into account the (reduced) density matrix elements for $X$, Equation~(\ref{eq4}).
In this way, we can only deal with reduced dynamics, which lack precise global information on
the coherence swapping between the two subsystems. Only the overall effects felt by $X$ are
described (e.g., the gradual suppression of interference features), which does not remove other
aspects, such as the non-crossing displayed by Bohmian trajectories, directly related to the
suppression of mutual coherence between the waves passing through each slit (alternative
mechanisms have to be incorporated in order to mimic this effect \cite{sanz:CPL:2009-2}).

To overcome the lacks and flaws involved in the above-mentioned procedure, particularly in those cases
where the focus lies on investigating how the coherence swapping takes place, it is also possible
to proceed by directly tackling the full problem through Equation~(\ref{eq1}), and, then, computing the
trajectories for both $X$ and $Y$, bearing in mind that each set of trajectories renders
a dynamical picture on the corresponding subspace, and, hence, any interpretation must rely on this.
In this case, though, it is not necessary to consider the density matrix approach (unless
the joint state describes a statistical mixture, which is not the case here); the usual wave
vector approach suffices for the purpose.
Therefore, if the joint probability amplitude in the position space is denoted as $\Psi(x,y|t)$,
then the corresponding equations of motion (the so-called guidance equations in Bohmian
mechanics) are
\begin{eqnarray}
 \dot{x}(x,y|t) & = & v^X (x,y|t) = \frac{1}{m}\ {\rm Re} \left[ \frac{\hat{p}_x \Psi(x,y|t)}{\Psi(x,y|t)} \right]
  = \frac{1}{m} \frac{\partial S(x,y|t)}{\partial x} ,
 \label{eq7} \\
 \dot{y}(x,y|t) & = & v^Y (x,y|t) = \frac{1}{m}\ {\rm Re} \left[ \frac{\hat{p}_y \Psi(x,y|t)}{\Psi(x,y|t)} \right]
  = \frac{1}{m} \frac{\partial S(x,y|t)}{\partial y} ,
 \label{eq8}
\end{eqnarray}
which arise from the transport {equation} 
\begin{equation}
 \dot{\bf r}(x,y|t) = {\bf v}(x,y|t) = \frac{1}{m}\ {\rm Re} \left[ \frac{\hat{\bf p} \Psi(x,y|t)}{\Psi(x,y|t)} \right]
  = \frac{1}{m} \frac{\partial S(x,y|t)}{\partial y} ,
 \label{eq9}
\end{equation}
where $\hat{\bf p} = -i\hbar \nabla$ denotes the usual momentum vector operator, with components
$(\hat{p}_x,\hat{p}_y)$ $ = -i\hbar (\partial/\partial x, \partial/\partial y)$, and
\begin{equation}
 S(x,y|t) = \frac{\hbar}{2i}\ \ln \left[ \frac{\Psi(x,y|t)}{\Psi^*(x,y|t)} \right]
 \label{eq10}
\end{equation}
is a field that specifies the local phase associated with the wave function $\Psi(x,y|t)$.\linebreak 
As can be noted, proceeding this way, the coupling between both subsystems becomes more apparent
through the dependence of both velocities on the two coordinates.
Moreover, also note that, in this manner, the information provided by the two velocity components
allows us to not only better understand the reduced dynamics displayed by $X$ later on, but also to better understand phenomena
such as quantum erasure \cite{walborn:PRA:2002}, or the so-called \mbox{coherence swapping
\cite{pati:Pramana:2001}.}


\subsection{Dynamical Behavior for Uncorrelated Systems}
\label{sec22}

The case of uncorrelated bipartite systems is trivial, because the quantum state of each
party evolves separately.
Yet, a brief analysis of the separate dynamics exhibited by one of the parties is important to
better understand the behavior of both the full uncorrelated system and, of course,
the entangled system.
Thus, consider that the joint state vector for the uncorrelated bipartite system is
\begin{equation}
 |\Psi(t)\rangle = |\Psi^X(t)\rangle \otimes |\Psi^Y(t)\rangle .
 \label{eq11}
\end{equation}
In the position representation, the corresponding wave function reads as
\begin{equation}
 \Psi(x,y|t) = \Psi^X(x,t) \Psi^Y(y,t) ,
 \label{eq12}
\end{equation}
which leads to independent probability densities and transverse velocity fields for each subsystem.
In other words, if these quantities are plotted in the corresponding subspaces ($X$ or $Y$), no
influence of one of the susbsystems on the other, and vice versa, is observed.
Therefore, if we consider a joint representation, we can easily identify the features that
belong to one or the other subsystem by simply inspecting what happens along the coordinate ($x$ or $y$)
that represents the corresponding subspace.

Below, we briefly analyze the main features associated with the dynamics exhibited by the time
evolution of both a single Gaussian wave packet and a coherent superposition of two Gaussian wave
packets, since they are used later to construct the bipartite states\linebreak  (see Section~\ref{sec23}).
Further details on the dynamical behaviors exhibited by these particular \textls[-15]{types of quantum states can
be found in \cite{sanz:JPA:2008,sanz:AJP:2012,sanz-bk-2}.
{
Physically, we here consider this particular} type of wave packet, because it facilitates direct comparison
with realistic physical scenarios of atom or molecular diffraction experiments, where short-range
interactions between diffracting mask and diffracted object lead to transmission amplitudes
describable in terms of Gaussian functions in a good approximation (e.g., see
\cite{sanz:AOP:2015}).
These Gaussian amplitudes are, thus, connected to the concept of the Gaussian slit introduced
by Feynman and Hibbs \cite{feynman-bk1,feynman-bk1-2010}, differing from the usual
semiclassical connotations associated with this type of wave packet.
Analogously, in the case of light, Gaussian beams can readily be associated with
coherent states produced by laser sources coupled to optical fibers, which also model in
a good approximation of real experiments (e.g., see \cite{sanz:PhysScr:2013}).}


\subsubsection{Single Gaussian Wave-Packet Dynamics}
\label{sec221}
In this section, as well as in Section~\ref{sec222}, we analyze the dynamics of a single system.
Consequently, to simplify notations, all unnecessary superscripts and subscripts are removed,
considering the $x$-coordinate to be generic.
Thus, consider that the wave function at $t=0$, $\Psi (x,0)$, has the general functional form of a
Gaussian wave packet with a phase (momentum) term
\vspace{-4pt}
\begin{equation}
 \mathcal{G}_0(x,0) = \left(\frac{1}{2\pi\sigma_0}\right)^{1/4}
  e^{-(x - x_0)^2/4\sigma_0^2 + ip_0 (x - x_0)/\hbar} ,
 \label{eq13}
\end{equation}
where $(x_0,p_0)$ specifies the phase--space position of its centroid and $\sigma_0$
its initial width.
In free space, the time-evolved form reads as \cite{sanz-bk-2}
\begin{equation}
 \mathcal{G}_0(x,t) = \left( \frac{1}{2\pi\tilde{\sigma}_t^2} \right)^{1/4}
  e^{- (x - x_t)^2/4\sigma_0\tilde{\sigma}_t + ip_0 (x - x_t)/\hbar
   + iE_0 t/\hbar} ,
 \label{eq14}
\end{equation}
which is also a Gaussian wave packet that propagates following the classical trajectory
$x_t = x_0 + v_0 t$, with $v_0 = p_0/m$ and $E_0 = p_0^2/2m$, and where its spatial dispersion
or spreading is described by the time-dependent complex function
\begin{equation}
 \tilde{\sigma}_t = \sigma_0 \left( 1 + \frac{i\hbar t}{2m\sigma_0^2} \right) .
 \label{eq15}
\end{equation}
Rewriting the spreading function (\ref{eq15}) in polar form, as
$\tilde{\sigma}_t = \sigma_t e^{i\varphi_t}$, we note
that the dispersion of the wave packet (\ref{eq14}) is given by the function
\begin{equation}
 \sigma_t =
  \sigma_0 \sqrt{ 1 + \left( \frac{\hbar t}{2m\sigma_0^2} \right)^2 } ,
 \label{eq16}
\end{equation}
while the phase factor that develops as the wave packet evolves in time is
\begin{equation}
 \varphi_t =
 \left( \tan \right)^{-1} \left( \frac{\hbar t}{2m\sigma_0^2} \right) ,
 \label{eq17}
\end{equation}
which can be related to the so-called geometric or Pancharatnam--Berry phase (or the Gouy phase
in the context of optical Gaussian beams \cite{sanz:ApplSci:2020}).
This latter term is responsible for the underlying dynamical behavior exhibited by the wave packet,
as can readily be noticed if the Gaussian wave packet (\ref{eq14}) is recast as
%
\begin{eqnarray}
 \mathcal{G}_0(x,t) & = & \left( \frac{1}{2\pi\sigma_t^2} \right)^{1/4}
  e^{-(x - x_t)^2 e^{-i\varphi_t}/4\sigma_0\sigma_t}
  e^{(i/\hbar)\left[ p_0 (x - x_t) + E_0 t - \varphi_t/2 \right] } \nonumber \\
  & = & \left( \frac{1}{2\pi\sigma_t^2} \right)^{1/4}
  e^{-(x - x_t)^2/4\sigma_t^2}
  e^{(i/\hbar) \left[ ((\hbar^2 t)/2m\sigma_0^2) (x - x_t)^2/4\sigma_t^2
   + p_0 (x - x_t) + E_0 t - \varphi_t/2 \right] } .
 \label{eq18}
\end{eqnarray}
This specific functional form for the wave packet makes the development of
time-dependent phase factors more apparent and how this leads, in general, to dispersive effects, while the translation
(described by the average momentum $p_0$ in this case) is only concerned with an intrinsically
classical (Newtonian-like) law of motion (here, a uniform motion).

In any of the standard quantum pictures, the presence of the above-mentioned phase factor
only manifests indirectly if the wave packet is made to interfere with itself (e.g.,~scattering
off of an impenetrable barrier) or in a coherent superposition with another wave packet (see
Section~\ref{sec222}, since it is not a quantum observable, unlike the probability density, which is
directly measurable with a detector, even on an event-by-event basis \cite{sanz:AnnFondLdB:2021}.
However, the hydrodynamical or Bohmian picture of quantum mechanics precisely highlights the role
played by the spatial variations of the phase \cite{sanz:AJP:2012,sanz:FrontPhys:2019}, and, hence,
renders a different perspective on this phase factor.
More specifically, such variations determine the local value of the transverse velocity field that
acts on the probability density, which manifests in the form of a quantum flux \cite{schiff-bk}.
This allows us to introduce the transport Equation (\ref{eq9}), which, in the case of the single
Gaussian wave packet (\ref{eq14}), acquires the specific functional form
\begin{equation}
 \dot{x}(x,t) = v(x,t) = v_0 + \frac{\hbar^2 t}{4m^2\sigma_0^2} \frac{(x - x_t)}{\sigma_t^2} .
 \label{eq19}
\end{equation}
This equation of motion contains information on both the translational motion, through the classical
velocity, $v_0$, and the dispersive effect, connected to the phase factor, $\varphi_t$.
As can be noted in Equation~(\ref{eq19}), although the transverse velocity field $v(x,t)$ has the
same value ($v_0$) at any position at $t=0$, at later times it becomes proportional to the
ratio $x/t$,\linebreak  which explains the linear asymptotic spreading exhibited by the wave packet.

The integration in time of the equation of motion (\ref{eq19}) is analytical and renders
\begin{equation}
 x(t) = x_t + \frac{\sigma_t}{\sigma_0} \left[ x(0) - x_0 \right] .
 \label{eq20}
\end{equation}
Following this equation, we find that, in the long-time limit, the Bohmian trajectories also
describe a classical-like (Newtonian) uniform motion,
\begin{equation}
 x(t) \sim v_\infty t ,
 \label{eq21}
\end{equation}
with the asymptotic constant velocity being
\begin{equation}
 v_\infty = v_0 + \frac{\hbar \left[ x(0) - x_0 \right]}{2m\sigma_0^2} .
 \label{eq22}
\end{equation}
This constant velocity, which only depends on how far from the centroid the Gaussian wave packet
trajectories are launched, is a direct consequence of the asymptotic linear behavior displayed by the transverse velocity field (\ref{eq19}). Note that Equation~(\ref{eq22}) corresponds to the long-time limit of Equation~(\ref{eq19}) when it is evaluated along a given trajectory $x(t)$:
\newpage
\begin{equation}
 v[x(t)] = v_0 + \frac{\hbar^2 t}{4m^2\sigma_0^2} \frac{[x(0) - x_0]}{\sigma_0\sigma_t} .
 \label{eq19b}
\end{equation}

\subsubsection{Young-Type Superposition Dynamics}
\label{sec222}

Young's interference experiment belongs to a class of quantum scenarios where the state vector can be
recast as a linear superposition ($S$) of different spatially separated alternatives.
In particular, we describe the full state of a particle diffracted by two slits as
\begin{equation}
 |\Psi_S (t)\rangle =  \frac{1}{\sqrt{2}} \left[ |\Psi_A (t)\rangle  + |\Psi_B (t) \rangle \right] ,
 \label{eq23}
\end{equation}
which physically represents a coherent superposition having an incident wave diffracted by
two slits, $A$ and $B$, spatially separated (ensuring orthogonality) and with the same amplitude
(probability).
In the position representation, we can represent this situation by means of the coherent superposition
of two Gaussian wave packets describing the diffracted wave coming up from each slit (e.g., the outcomes
in an experiment such as the one reported in \cite{kocsis:Science:2011}):
\begin{equation}
 \Psi_S (x,t) =  \mathcal{N} \left[ \mathcal{G}_A (x,t) + \mathcal{G}_B (x,t) \right] ,
 \label{eq24}
\end{equation}
where
\begin{eqnarray}
 \mathcal{G}_A (x,t) & = & \left( \frac{1}{2\pi\tilde{\sigma}_t^2} \right)^{1/4}
  e^{- (x - x_{A,t})^2/4\sigma_0\tilde{\sigma}_t
   + ip_A (x - x_{A,t})/\hbar + i E_A t/\hbar } ,
 \label{eq25} \\
 \mathcal{G}_B (x,t) & = & \left( \frac{1}{2\pi\tilde{\sigma}_t^2} \right)^{1/4}
  e^{- (x - x_{B,t})^2/4\sigma_0\tilde{\sigma}_t
   + ip_B (x - x_{B,t})/\hbar + i E_B t/\hbar } ,
 \label{eq26}
\end{eqnarray}
where $X_{A/B,t}$ has the same meaning as $X_t$ in the previous case, but with $A$ and $B$
denoting the corresponding centroid position, namely, $x_{A,0} = d/2 = - x_{B,0}$, with
$d$ being the separation between both centroids (which is the same as the center-to-center
distance between the two slits).
Regarding the respective momenta, for simplicity, we consider no transverse drift, as this is not relevant to the
discussion here,
so $p_A = p_B = 0$\linebreak  (and, hence, $E_A = E_B = 0$), 
On the other hand, the two Gaussian wave packets might not have a vanishing overlapping, since
\begin{equation}
 \mathcal{N} = \frac{1}{\sqrt{2 \left(1 + e^{-d^2/8\sigma_0^2}\right)}} ,
 \label{eq27}
\end{equation}
thus, invalidating the condition of mutual orthogonality between the states representing the diffraction
through the slits.
To overcome this inconvenience, from now on we assume the condition $d/2 \gg \sigma_0$, so that
$\mathcal{N} \approx 1/\sqrt{2}$.

From the density matrix coefficients associated with the wave function (\ref{eq24}),
\begin{equation}
 \rho(x,x'|t)
 \approx \frac{1}{2} \Big[ \mathcal{G}_A(x,t)\mathcal{G}_A^*(x',t)
    + \mathcal{G}_B(x,t)\mathcal{G}_B^*(x',t)
    + \mathcal{G}_A(x,t)\mathcal{G}_B^*(x',t)
    + \mathcal{G}_B(x,t)\mathcal{G}_A^*(x',t) \Big] ,
 \label{eq28}
\end{equation}
we readily obtain the probability density, which reads as
\begin{equation} 
 \rho(x,t) \approx \frac{1}{2}\left( \frac{1}{2\pi\sigma_t^2} \right)^{1/2}
  \left[ e^{-(x - d/2)^2/2\sigma_t^2} + e^{-(x + d/2)^2/2\sigma_t^2}
   + 2 e^{-[x^2 + (d/2)^2]/2\sigma_t^2} \cos k_t x \right] ,
 \label{eq29}
\end{equation}
where
\begin{equation}
 k_t = \frac{\hbar t d}{4m\sigma_0^2\sigma_t^2} .
 \label{eq30}
\end{equation}
For relatively short times, compared to $\tau = 2m\sigma_0^2/\hbar$ (i.e., $t \ll \tau$),
we have $\sigma_t \approx \sigma_0$,\linebreak  and Equation~(\ref{eq41}) essentially consists of two
separated Gaussian wave packets centered around $x_A$ and $x_B$:
\begin{equation} 
 \rho(x,t) \approx \frac{1}{2}\left( \frac{1}{2\pi\sigma_0^2} \right)^{1/2}
  \left[ e^{-(x - d/2)^2/2\sigma_0^2} + e^{-(x + d/2)^2/2\sigma_0^2} \right] .
 \label{eq31}
\end{equation}
On the other hand, for $t \gg \tau$, the width of each wave packet increases linearly with time,
since $\sigma_t \approx \hbar t/2m\sigma_0$ (see Section~\ref{sec221}), and a Young-type
interference pattern arises with a Gaussian envelope:
\begin{equation} 
 \rho(x,t) \approx 2 \left( \frac{2 m^2 \sigma_0^2}{\pi \hbar^2 t^2} \right)^{1/2}
   e^{-2m^2\sigma_0^2 x^2/\hbar^2 t^2} \cos^2 (k_\infty x/2) ,
 \label{eq32}
\end{equation}
with
\begin{equation}
 k_\infty = \frac{md}{\hbar t} .
\label{eq33}
\end{equation}
As is inferred from (\ref{eq32}), at a given time, the condition to observe
an interference minimum~is
\begin{equation}
 x_n = \left( n + \frac{1}{2} \right) \frac{2\pi\hbar t}{md} ,
 \label{eq34}
\end{equation}
where $n = 0, \pm 1, \pm 2, \ldots$.
Therefore, the size of the interference fringes (distance between two consecutive minima)
increase linearly with time, as
\begin{equation}
 \Delta x = \frac{2\pi\hbar t}{md} ,
 \label{eq35}
\end{equation}
while the number of interference fringes within the region covered by the Gaussian\linebreak  envelope
increase rapidly as $d$ decreases.

In order to understand the appearance of fringes from the dynamical point of
view rendered by the phase variations of the wave function (\ref{eq24}), let us
now examine the \linebreak expression for $v(x,t)$, which is readily obtained after
substituting this wave function into Equation~(\ref{eq7}) in the factorizable
case:
\begin{eqnarray}
 \dot{x}(x,t) & \approx &
 \frac{1}{2\rho(x,t)} \left( \frac{1}{2\pi\sigma_t^2} \right)^{1/2}
  \frac{\hbar^2 t}{4m^2\sigma_0^2 \sigma_t^2}
 \bigg\{ (x - d/2) e^{-(x - d/2)^2/2\sigma_t^2}
  + (x + d/2) e^{-(x + d/2)^2/2\sigma_t^2} \nonumber \\
  & &
  + 2 x e^{-[x^2 + (d/2)^2]/2\sigma_t^2} \cos k_t x
  + \frac{2m\sigma_0^2 x_0}{\hbar t}\
  e^{-[x^2 + (d/2)^2]/2\sigma_t^2} \sin k_t x \bigg\} .
 \label{eq36}
\end{eqnarray}
This highly nonlinear expression already gives us an idea of the complexity of the dynamics.
Yet, some conclusions can be extracted from it even without solving it numerically.
Since this general expression was already discussed in detail in \cite{sanz:JPA:2008},
here we focus on the above two limits.
Note that this expression cancels out at $x=0$ (which is at the origin of the so-called
Bohmian non-crossing rule \cite{sanz:JPA:2008}), although it remains finite on either
side of this boundary.
Thus, in the short-time limit, although there is a non-negligible contribution, the density
mainly concentrates around the slit $A$ or the slit $B$ (i.e., around $x_A = d/2$ or $x_B = -d/2$).
Accordingly, the trajectories launched from a vicinity of these points
describes a motion analogous to that exhibited by the trajectories associated with a free
Gaussian wave packet, because Equation~(\ref{eq36}) approximates
\begin{equation}
 \dot{x}_c(x,t) \approx \frac{\hbar^2 t}{4m^2\sigma_0^4}\ (x - x_0) ,
 \label{eq37}
\end{equation}
with $c = A, B$.
Thus, the larger $d$, the more the trajectories resemble those
for a free Gaussian (centered either on site $A$ or site $B$).

In the long-time limit, though, we find rather different behavior with respect to the single-Gaussian
case, as Equation~(\ref{eq36}) approximates
\begin{equation}
 \dot{x}(x,t) \approx \frac{x}{t}
  \left[ 1 - \frac{m\sigma_0^2 d}{2\hbar t x}\ \tan (k_\infty x/2) \right] .
 \label{eq38}
\end{equation}
Accordingly, at a given time, we observe a nearly linear increase with $x$,
except at the regions around an interference minimum, specified by the condition (\ref{eq34}),
where the velocity undergoes a sudden change because the argument of the tangent equals $\pi/2$.
For positive $x$, these sudden changes in the velocity are in the form of
negative dips or spikes, while for negative $x$ they have the form of
positive spikes.
As a consequence, at relatively large $t$, trajectories tend to mainly accumulate
within the region determined by two consecutive neighboring spikes, where they
display a nearly uniform motion, with a (quantized) mean value for the momentum given,
in a good approximation, by the expression
\begin{equation}
 \hbar \kappa_n = m\ v(\bar{x},t)
 = m\ v[(x_{n+1/2} + x_{n-1/2})/2,t]
 \approx \frac{2\pi\hbar}{d}\ n ,
 \label{eq39}
\end{equation}
with $n = 0, \pm 1, \pm 2, \ldots$, and where $x_{n+1/2}$ and $x_{n-1/2}$ denote
the position of the two spikes surrounding the region with momentum $\kappa_n$.
This momentum value, quantized in terms of the unit $2\pi\hbar/d$, determines the
long-time position (on average) of interference maxima: $x_n \approx \hbar \kappa_n t/m$.
In this way, swarms of trajectories with nearly the same constant velocity
constitute the dynamical consequence of the organized phase variations arising
after the overlapping of the two mutually (phase) coherent Gaussian wave packets.
The probability density, statistically specified in terms of these swarms of
trajectories, thus tends to accumulate in the regions where the transverse velocity
field is nearly constant, between any two consecutive velocity ``spikes'', while it
is expelled by means of a rather fast flow when it is in the vicinity of these ``spikes'',
always from the inner regions to the outer ones.
This is analogous to the accumulation of sediments on the sandbanks of a river,
where the watercourse is slow, while they are abruptly expelled in the rapids
because of the fast streams.


\subsection{Entangled Bipartite Systems}
\label{sec23}

Let us now focus on the case of a Bell-type maximally entangled ($E$) state:
\begin{equation}
 |\Psi_E (t)\rangle = \frac{1}{\sqrt{2}} \left[ |\psi_A^X (t)\rangle |\psi_B^Y (t)\rangle + |\psi_B^X (t)\rangle |\Psi_A^Y (t)\rangle\right] ,
 \label{eq40}
\end{equation}
which describes a situation with two possible slits, $A$ and $B$, and two
correlated identical particles, $X$ and $Y$, that can get diffracted through
either of the slits, but not both,\linebreak  as in the standard superposition case seen in the
previous section.
More specifically, unlike the (factorizable) bipartite vector state for both particles
undergoing diffraction through both slits,
\begin{eqnarray}
 |\Psi_S (t)\rangle & = & |\Psi_S^X (t)\rangle \otimes |\Psi_S^Y (t)\rangle \nonumber \\
 & = &
 \frac{1}{2} \left[ |\psi_A^X (t)\rangle |\psi_A^Y (t)\rangle
  + |\psi_A^X (t)\rangle |\psi_B^Y (t)\rangle + |\psi_B^X (t)\rangle |\psi_A^Y (t)\rangle
  + |\psi_B^X (t)\rangle |\psi_B^Y (t)\rangle \right] ,
 \label{eq41}
\end{eqnarray}
the entangled state (\ref{eq40}) lacks the possibility to have both particles
diffracted through the same slit {[}first and fourth terms in (\ref{eq41}){],}
thus, describing a highly delocalized two-particle system.
To further investigate the associated dynamics, let us consider the wave
function in the position representation for (\ref{eq40}),
\begin{equation}
 \Psi_E (x,y|t) =  \mathcal{N}_E \left[ \mathcal{G}_A (x,t) \mathcal{G}_B (y,t)
   + \mathcal{G}_B (x,t) \mathcal{G}_A (y,t) \right] ,
 \label{eq42}
\end{equation}
where the wave packets are as given by (\ref{eq25}) and (\ref{eq26}),
although substituting $x$ by $y$ wherever it corresponds, and the normalizing
prefactor is
\begin{equation}
 \mathcal{N}_E = \frac{1}{\sqrt{2 \left( 1 + e^{- d^2/4\sigma_0^2} \right)}} .
 \label{eq43}
\end{equation}
As before, the choice of $d$ makes $\mathcal{N}_E \approx 1/\sqrt{2}$.

From the above considerations, we obtain the expression for the density  {matrix \linebreak elements,}
\begin{eqnarray}
 \rho(x,y;x',y'|t)
 & \approx & \frac{1}{2} \Big[
  \mathcal{G}_A(x,t) \mathcal{G}_A^*(x',t)
  \mathcal{G}_B(y,t) \mathcal{G}_B^*(y',t)
  + \mathcal{G}_B(x,t) \mathcal{G}_B^*(x',t)
    \mathcal{G}_A(y,t) \mathcal{G}_A^*(y',t) \nonumber \\ & &
  + \mathcal{G}_A(x,t) \mathcal{G}_B^*(x',t)
    \mathcal{G}_B(y,t) \mathcal{G}_A^*(y',t)
  + \mathcal{G}_B(x,t) \mathcal{G}_A^*(x',t)
    \mathcal{G}_A(y,t) \mathcal{G}_B^*(y',t) \Big] ,
 \label{e44}
\end{eqnarray}
from which the diagonal elements provide us with the bipartite probability density,
\begin{eqnarray}
 \rho(x,y|t) & \approx & \frac{1}{2} \left( \frac{1}{2\pi\sigma_t^2} \right)
 \bigg\{ e^{- (x - d/2)^2/2\sigma_t^2 - (y + d/2)^2/2\sigma_t^2}
  + e^{- (x + d/2)^2/2\sigma_t^2 - (y - d/2)^2/2\sigma_t^2}
  \nonumber \\ & &
  + 2 e^{- [x^2 + y^2 + 2(d/2)^2]/2\sigma_t^2}
    \cos \left[ k_t (x - y) \right] \bigg\} .
 \label{eq45}
\end{eqnarray}
Let us now proceed as before and investigate the short-time and long-time behaviors
exhibited by this probability density.
In the short-time limit ($t \ll \tau$), it basically describes two highly
delocalized systems with maximal uncertainty about both \cite{sanz:ENTROPY:2022}:
\begin{equation}
 \rho(x,y|t) \approx \frac{1}{2} \left( \frac{1}{2\pi\sigma_0^2} \right)
 \bigg[ e^{- (x - d/2)^2/2\sigma_0^2 - (y + d/2)^2/2\sigma_0^2}
  + e^{- (x + d/2)^2/2\sigma_0^2 - (y - d/2)^2/2\sigma_0^2} \bigg] .
 \label{eq46}
\end{equation}
On the other hand, in the long-time limit ($t \gg \tau$), we have
\begin{equation}
 \rho(x,y|t) \approx 2 \left( \frac{2m^2\sigma_0^2}{\pi\hbar^2 t^2} \right)
  e^{- (x^2 + y^2)/2\sigma_t^2}
 \cos^2 \left[ k_\infty (x - y)/2 \right] .
 \label{eq47}
\end{equation}
According to this expression, the density distribution exhibits full fringe
visibility along the direction $y=-x$, since
\begin{equation}
 \rho_\infty(x,-x|t) \approx 2 \left( \frac{2m^2\sigma_0^2}{\pi\hbar^2 t^2} \right)
  e^{- 4m^2\sigma_0^2 x^2/\hbar^2 t^2} \cos^2 \left( k_\infty x \right) ,
 \label{eq45bis}
\end{equation}
while it is totally suppressed along the $y=x$, where
\begin{equation}
 \rho_\infty(x,x|t) \approx 2 \left( \frac{2m^2\sigma_0^2}{\pi\hbar^2 t^2} \right)
  e^{- 4m^2\sigma_0^2 x^2/\hbar^2 t^2} .
 \label{eq45biss}
\end{equation}

The presence of the oscillatory term in Equation~(\ref{eq45}) can be interpreted as a signature
of the coherence exchange, due to correlations between the $X$ and $Y$ subsystems, which is
maximal, as we have seen, along the direction that joins them directly. It disappears
gradually as the perpendicular direction is approached (as should be expected, since there
was no probability to find both subsystems at the same position initially, i.e., around
$x=y=\pm d/2$).
This strong correlation has a direct implication when we investigate the behavior of the
subsystem $X$.
Tracing over the coordinates for the $Y$ subsystem in~(\ref{e44}), we obtain the following
expression for the reduced density matrix elements of the \mbox{subsystem $X$:}
\begin{equation}
 \tilde{\rho}(x,x'|t)
 \approx \frac{1}{2} \bigg[
    \mathcal{G}_A(x,t) \mathcal{G}_A^*(x',t)
  + \mathcal{G}_B(x,t) \mathcal{G}_B^*(x',t)
  + \Lambda_{AB}^* \mathcal{G}_A(x,t) \mathcal{G}_B^*(x',t)
  + \Lambda_{AB} \mathcal{G}_B(x,t) \mathcal{G}_A^*(x',t) \bigg] ,
 \label{eq48}
\end{equation}
where
\begin{equation}
 \Lambda_{AB} = \int \mathcal{G}_A(y,t) \mathcal{G}_B^*(y,t) dy
  = e^{- d^2/8\sigma_0^2} .
 \label{eq49}
\end{equation}
Therefore, the reduced probability density reads as
\begin{equation}
 \tilde{\rho}(x,t)
 \approx \frac{1}{2} \left( \frac{1}{2\pi\sigma_t^2} \right)^{1/2}
  \bigg[ e^{- (x - d/2)^2/2\sigma_t^2} + e^{- (x + d/2)^2/2\sigma_t^2}
  + 2\Lambda_{AB} e^{- [x^2 + (d/2)^2]/2\sigma_t^2} \cos k_t x \bigg] ,
 \label{eq50}
\end{equation}
where the interference term is partially, or even totally, suppressed by the
prefactor $\Lambda_{AB}$, depending on the balance between the width of the wave
packets and their distances.
Since we typically consider $d/2$ to be larger than $\sigma_0$ to avoid the initial
overlapping of the wave packets, it is clear that (\ref{eq50}) represents in
both the short-time and the long-time limits an incoherent sum of two independent
wave packets.
It is precisely because of this incoherence effect in the $X$-subspace, in the long-time limit, that the density distribution is basically described by a seemingly
single Gaussian distribution:
\begin{equation}
 \tilde{\rho}(x,t)
 \approx \left( \frac{2m\sigma_0^2}{\pi\hbar t} \right)^{1/2}
  e^{-x^2/2\sigma_t^2} .
 \label{eq51}
\end{equation}
This arises as a consequence of the screening produced by the correlation with
the $Y$ subsystem, which leads to a statistical mixture, without phase relations,
in the $X$ subsystem, according to Schr\"odinger's picture of entanglement \cite{schrodinger:ProcCamPS:1936}.

The above behavior is what we commonly regard as decoherence \cite{zurek:PhysToday:1991,zurek:PhysToday-rev:2003,omnes:RMP:1992,schlosshauer:RMP:2004}, that is,\linebreak  the loss of phase correlations
by interaction with other quantum systems.
However, even after having lost such phase correlations and observing seemingly classical
behavior for the $X$ subsystem, the point is that, if we analyze the corresponding velocity
fields, the correlations survives due to the nonlinear nature of these fields, both
in the full space and also in the reduced one.
More specifically, if both subsystems are considered, we obtain the following equations of
{motion:} 
\begin{eqnarray}
 \dot{x}(x,y|t) & \approx &
 \frac{1}{2\rho(x,y|t)} \left( \frac{1}{2\pi\sigma_t^2} \right)
  \frac{\hbar^2 t}{4m^2\sigma_0^2 \sigma_t^2} \nonumber \\
  & & \times
  \bigg\{ (x - d/2) e^{-(x - d/2)^2/2\sigma_t^2 - (y + d/2)^2/2\sigma_t^2}
  + (x + d/2) e^{-(x + d/2)^2/2\sigma_t^2 - (y - d/2)^2/2\sigma_t^2} \nonumber \\
  & &
  + 2 x e^{-[x^2 + y^2 + 2(d/2)^2]/2\sigma_t^2} \cos [k_t (x-y)]
  + \frac{m\sigma_0^2 d}{\hbar t}\
  e^{-[x^2 + y^2 + 2(d/2)^2]/2\sigma_t^2} \sin [k_t (x-y)] \bigg\} ,
  \nonumber \\ & &
 \label{eq52} \\
 \dot{y}(x,y|t) & \approx &
 \frac{1}{2\rho(x,y|t)} \left( \frac{1}{2\pi\sigma_t^2} \right)
  \frac{\hbar^2 t}{4m^2\sigma_0^2 \sigma_t^2} \nonumber \\
  & & \times
  \bigg\{ (y + d/2) e^{-(x - d/2)^2/2\sigma_t^2 - (y + d/2)^2/2\sigma_t^2}
  + (y - d/2) e^{-(x + d/2)^2/2\sigma_t^2 - (y - d/2)^2/2\sigma_t^2} \nonumber \\
  & &
  + 2 y e^{-[x^2 + y^2 + 2(d/2)^2]/2\sigma_t^2} \cos [k_t (x-y)]
  - \frac{m\sigma_0^2 d}{\hbar t}\
  e^{-[x^2 + y^2 + 2(d/2)^2]/2\sigma_t^2} \sin [k_t (x-y)] \bigg\} , 
  \nonumber \\ & &
 \label{eq53}
\end{eqnarray}
which provide us with relevant information on the strong correlation (and its effects)
between the two subsystems through their nonseparability.
According to these equations, the evolution of $X$ is strongly connected to the evolution
of $Y$, and vice versa.

On the other hand, in the reduced case, although the explicit presence of the\linebreak  $y$-coordinate
has been removed, its effects are still felt through the prefactor $\Lambda_{AB}$,\linebreak  and we have
\begin{eqnarray}
 \dot{\tilde{x}}(x,t) & \approx &
 \frac{1}{2\tilde{\rho}(x,t)} \left( \frac{1}{2\pi\sigma_t^2} \right)^{1/2}
  \frac{\hbar^2 t}{4m^2\sigma_0^2 \sigma_t^2}
  \bigg\{ (x - d/2) e^{-(x - d/2)^2/2\sigma_t^2}
  + (x + d/2) e^{-(x + d/2)^2/2\sigma_t^2} \nonumber \\
  & &
  + 2 x \Lambda_{AB} e^{-[x^2 + (d/2)^2]/2\sigma_t^2} \cos k_t x
  + \Lambda_{AB} \frac{m\sigma_0^2 d}{\hbar t}\
  e^{-[x^2 + (d/2)^2]/2\sigma_t^2} \sin k_t x \bigg\},
 \label{eq54}
\end{eqnarray}
which, given the value chosen for $d$, can be simplified to
\begin{equation}
 \dot{\tilde{x}}(x,t) \approx
  \left( \frac{\hbar^2 t}{4m^2\sigma_0^2 \sigma_t^2} \right)
  \bigg\{
  \frac{(x - d/2) e^{-(x - d/2)^2/2\sigma_t^2}
  + (x + d/2) e^{-(x + d/2)^2/2\sigma_t^2}}{e^{-(x - d/2)^2/2\sigma_t^2}
  + e^{-(x + d/2)^2/2\sigma_t^2}} \bigg\} .
 \label{eq55}
\end{equation}
Observe that, in the short-time limit, this expression renders two swarms of trajectories
that seem to be independent of one another, each one being associated with a single wave packet.
On the other hand, in the long-time limit, we have a nearly homogeneous
distribution of trajectories that seem to be reproducing the behavior of a single
motionless ($v_0=0$) wave packet centered at $x=0$, with their effective equation of
motion being
\begin{equation}
 \dot{\tilde{x}}(x,t) \approx
  \left( \frac{\hbar^2 t x}{4m^2\sigma_0^2 \sigma_t^2} \right) .
 \label{eq56}
\end{equation}

Although we have seen that the full and the reduced dynamics are very different from the
point of view of the flux (trajectories), the former still has to provide limiting
behaviors analogous to those exhibited by the latter.
In other words, although the full dynamics does not need to preserve the
Bohmian non-crossing property on the corresponding subspace, because it only remains
valid in the full-dimensional space, the trajectories have to show the same asymptotic
behavior in both cases (i.e., for $t \gg \tau$).


\section{Results and Discussion}
\label{sec3}


\subsection{Single-System Dynamical Behaviors}
\label{sec31}

\begin{figure}[!t]
 \centering
\includegraphics[width=\textwidth]{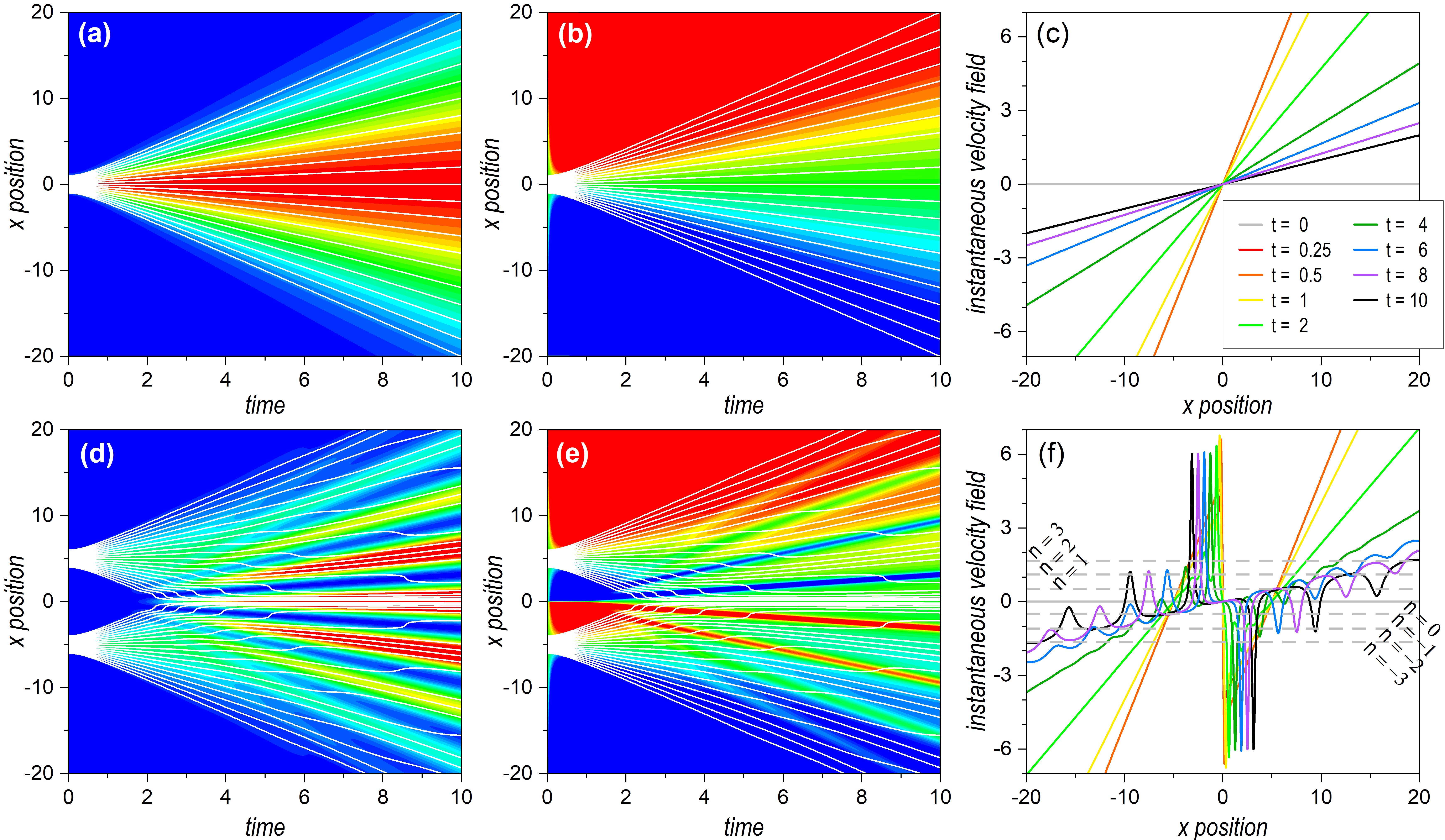}
 \caption{\label{Fig1}
 {Time} evolution of a single Gaussian wave packet centered at $x_0 = 0$ (\textbf{{upper}} \textbf{row} panels) and
  a coherent superposition of two Gaussian wave packets with $d = 10$ (\textbf{lower} \textbf{row} panels).\linebreak 
  (\textbf{a},\textbf{d}) Contour plot of the probability density.
  {In the color code, minimum density (zero) is denoted with blue
  and maximum with red.
  For a better visualization, in the case of the superposition state, we consider a truncation to about two thirds of
  the maximum.}
  (\textbf{b},\textbf{e}) Contour plot of the transverse velocity field (\ref{eq19}). For a better visualization, both the maximum (positive) and minimum (negative) values of the velocity fields were truncated to 1.5 and $-1.5$, respectively. In the color code, these values are denoted with red and blue, respectively, and greenish hues represent zero velocity  values and around this.  (\textbf{c},\textbf{f}) Transverse velocity field in terms of the $x$-coordinate at specific times  [see color legend in panel (\textbf{c}); since the yellow line (for $t=1$) overlaps the red one (for $t=0.25$), only the former can be seen in both panels (\textbf{c}) and (\textbf{f})].  To illustrate the dynamics generated by the velocity field, the Gaussian wave packets in each  contour plot are covered by a set of 21 Bohmian trajectories (white solid lines) with equidistant  initial conditions.  {In the two cases, the initial conditions were chosen within the interval $x_c \pm \Delta x$,  with $\Delta x = 1$, and where $x_c = x_0 = 0$ for the single Gaussian, and $x_c = x_A, x_B$, with
  $x_A = - x_B = 5$, for the superposition state.}  The horizontal dashed gray lines in panel (\textbf{f}) indicate the quantized value for the average momentum,  $\hbar \kappa_n$, determined according to Equation~(\ref{eq39}),\linebreak  from $n=0, \pm 1, \pm 2$, and $\pm 3$.  The numerical values used in the simulations were: $\sigma_0 = 0.5$, $p_0 = 0$, $m=1$, and $\hbar = 1$.  See text for further details.}
\end{figure}

In order to better understand the behaviors exhibited by the separable and entangled bipartite systems,
let us first briefly revisit the main dynamical aspects of single localized and delocalized systems,
respectively, described by the wave functions (\ref{eq14}) and (\ref{eq24}),\linebreak  and represented in the upper
and lower rows of Figure~\ref{Fig1}.
{For the single Gaussian wave packet (\ref{eq14}) (see Section~\ref{sec221}), the time evolution
of the corresponding probability density is represented in the form of a contour plot
in Figure~\ref{Fig1}a. As is seen, it exhibits} a slow dispersion initially until it undergoes a fast increase
around $t \approx \tau$ (with the parameters used in the simulations, $\tau = 0.5$), which
eventually leads to a linearly increasing dispersion regime. {The associated quantum flux is illustrated with the depiction of 21 Bohmian
trajectories, which follow Equation~(\ref{eq20}) and have equidistant initial positions.
We clearly} observe a transition from a swarm of nearly parallel
trajectories to a linear separation among them, with increasingly faster speeds as the initial condition
gets further away from the center of the distribution.

The dynamics described by the probability density are a manifestation of the underlying flux,
{ruled by the local variations of the phase term that appears in the wave function along its time evolution, as can clearly be seen in Equation~(\ref{eq18}).}
These phase variations, as mentioned before, generate an underlying transverse
{velocity field, Equation~(\ref{eq19}), which gradually changes in time, as can be observed in Figure~\ref{Fig1}b in the form of a contour plot, and in Figure~\ref{Fig1}c in the
form of cross sections at specific times.}
As seen in Figure~\ref{Fig1}c, at a given time, this velocity field increases linearly with $x$,
pivoting around $x=0$; thus, provoking trajectories with $x(0) > 0$ to acquire positive
outwards momentum and trajectories with $x(0) < 0$ acquiring negative outwards momentum.
This eventually leads to the spread of the whole swarm as time increases.
Indeed, as is also seen in Figure~\ref{Fig1}c, there is an almost linear increase of the slope of the
velocity field from zero, until it reaches a maximum value at $t = \tau$ {[}as can be inferred from
Equation~(\ref{eq19}){];} then, the slope starts decreasing again, with nearly $t^{-1}$, until it approaches
zero again.

\begin{figure}[!t]
 \centering
 \includegraphics[width=0.4\textwidth]{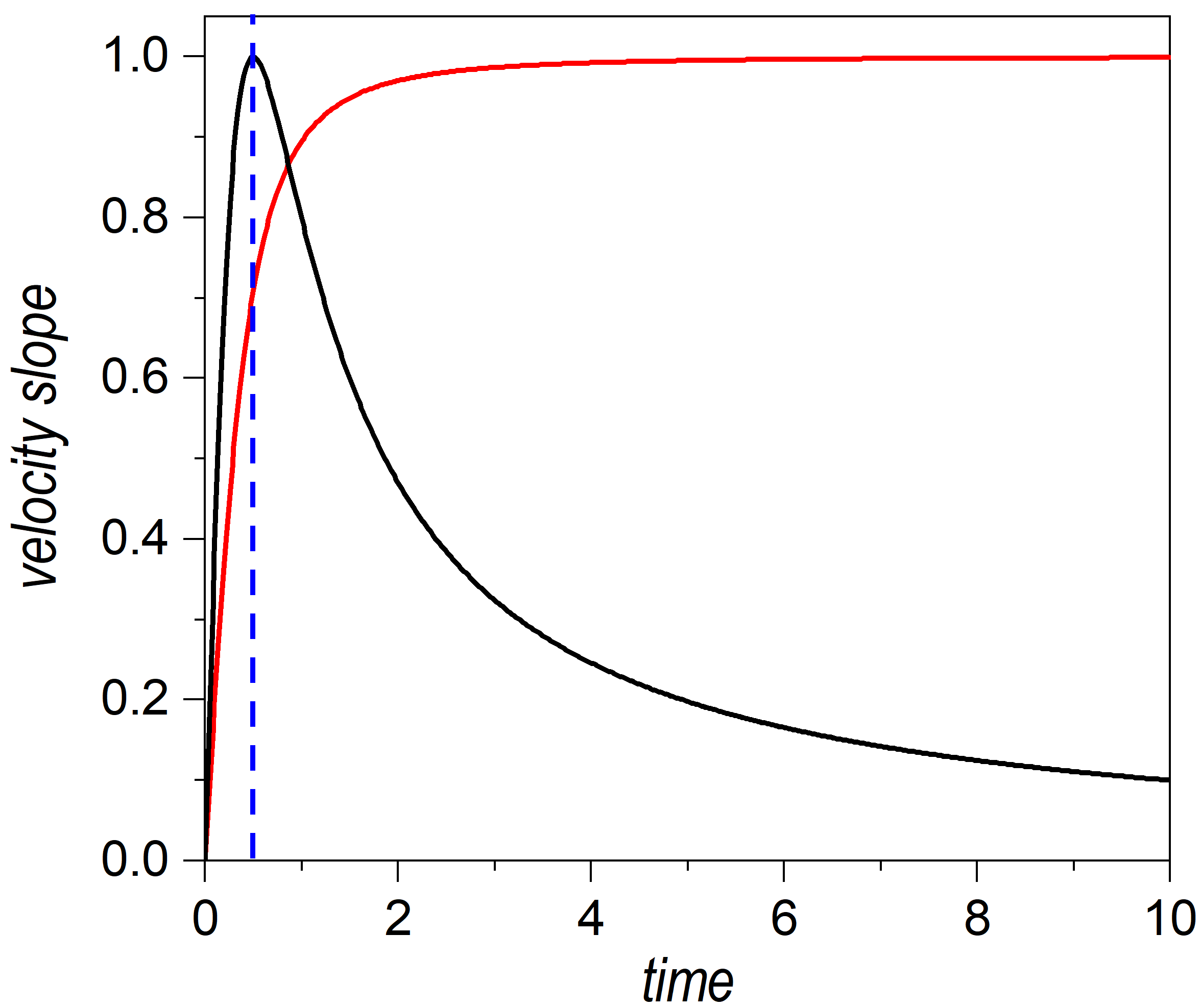}
 \caption{\label{Fig2}
  Time evolution of the time-dependent diffusive prefactors of Equations~(\ref{eq19}) and (\ref{eq19b}),\linebreak  which
  rule out, respectively, the slope of the velocity field $v(x,t)$ at a given time (solid black line) and
  the separation rate $v[x(t)]$ of the trajectories with respect to the centroid of the wave packet
  (solid red line).
  Although both slopes increase linearly at short times, in the first case a maximum is reached
  at $t=\tau$ (this characteristic time scale is denoted with the dashed vertical blue line), and then
  it starts decreasing as $t^{-1}$.
  In the second case, the slope keeps increases until it reaches an asymptotic constant value,
  which corresponds to the asymptotic value of the dispersion rate of a Gaussian wave packet,
  $d\sigma_t/dt \approx \hbar/2m\sigma_0$ (this is the so-called spreading velocity, as {introduced} in \cite{sanz:JPA:2008}).
  The numerical values considered here were: $\sigma_0 = 0.5$, $p_0 = 0$, $m=1$, and $\hbar = 1$.}
\end{figure}

Such behaviors become more apparent by inspecting the diffusive time-dependent prefactors in
Equations~(\ref{eq19}) and (\ref{eq19b}), for $v(x,t)$ and $v[x(t)]$, respectively, which are
displayed, in Figure~\ref{Fig2}, as functions of time.
In the case of the overall velocity field (black line), and taking into account the above-mentioned transition at $t = \tau$,
we effectively appreciate two regimes, namely, a nearly uniform acceleration for $t<\tau$
and, then, an asymptotic fall with $t^{-1}$ for $t>\tau$.
On the other hand, if we consider the velocity along a given trajectory (red line),
the first stage is characterized by nearly uniform acceleration, which provokes fast
diffusion of the trajectories, and, then, beyond $t = \tau$, the trajectories acquire a
constant speed, proper for uniform motion, as in classical mechanics.

The behaviors that explain the dynamics of a single Gaussian wave packet also hold for a coherent
superposition of two of them, like the one described by Equation~(\ref{eq24}), provided they are far enough from each other, as is
seen at early stages in Figure~\ref{Fig1}d,e, compared to Figure~\ref{Fig1}a,b, respectively.
There is a major difference, though, between the picture offered by the probability density here
and the plot of the velocity field. Although the velocity field is zero at $t=0$, because of the
existence of phase coherence, since the very early stages a sharp spatial division arises, from a
dynamical point of view, with two independent velocity fields that do not overlap, but that undergo
a sharp shear along the symmetry line $x=0$.
This behavior is observed until $t \approx 2$, when the dispersion of both wave packets is enough to provoke
important overlapping between them, analogous to that exhibited by the single wave packet,
although with the slope of the velocity field pivoting around the respective centroids.
This behavior is better appreciated with the aid of the corresponding trajectories, which
show the specific flux evolution at a more local level.
These trajectories, superimposed to the density plots, are obtained after numerical integration
of Equation~(\ref{eq36}) with initial conditions equally distributed within the regions covered
by each wave packet.
As can be seen, the two swarms of trajectories show a seemingly independent evolution (diffraction) at short times.

At later times, an increasingly prominent overlapping of the two wave packets leads to the appearance
of interference traits in the probability density, which are associated with the appearance of regions
with relatively constant velocity, interrupted by regions wherein the velocity undergoes sudden changes.
The first regions, more stable from a dynamical point of view, enable the accumulation of trajectories
along them with nearly the same velocity (on average), which explains the interference maxima.
On the other hand, the latter imprint an important shake on trajectories passing by too close to them,
which leads them to quickly change from one stable region to the neighboring one, in an inwards
staggered motion.
This promotion of trajectories from the outer regions of the wave function to the innermost ones
eventually leads to a higher population of the corresponding (innermost) interference maxima with
respect to the more marginal ones.

Another interesting feature in the long-time regime is that, from two spatially divided dynamical
regions, the velocity tends to exhibit a single average slope (interrupted by a series of sudden
kicks).
However, although this average slope seems to slowly decrease to zero, as we might infer from
Figure~\ref{Fig1}f, what actually happens is that it converges to an almost constant slope, in
compliance with the asymptotic value given by Equation~(\ref{eq38}), such that each stable section
(between any two kicks) remains at the same height with respect to the central value $v(0,t) = 0$.
The average value for the height reached by each section is given by Equation~(\ref{eq39}), as is
indicated by the corresponding horizontal dashed lines.
This behavior was actually observed in the experiments with photons performed by Kocsis et al.~\cite{kocsis:Science:2011} more than a decade ago.
With the aid of weak measurements, it was possible to measure the transverse momentum, which
is a quantity precisely proportional to the Bohmian instantaneous velocity field (although
the latter refers to massive particles and, in the case of photons, is related to the
transverse component of the wave vector under paraxial conditions
\cite{sanz:AnnPhysPhoton:2010,sanz:EPN:2013}).


\subsection{Bipartite-System Dynamical Behaviors}

\begin{figure}[!t]
\centering
\includegraphics[width=\textwidth]{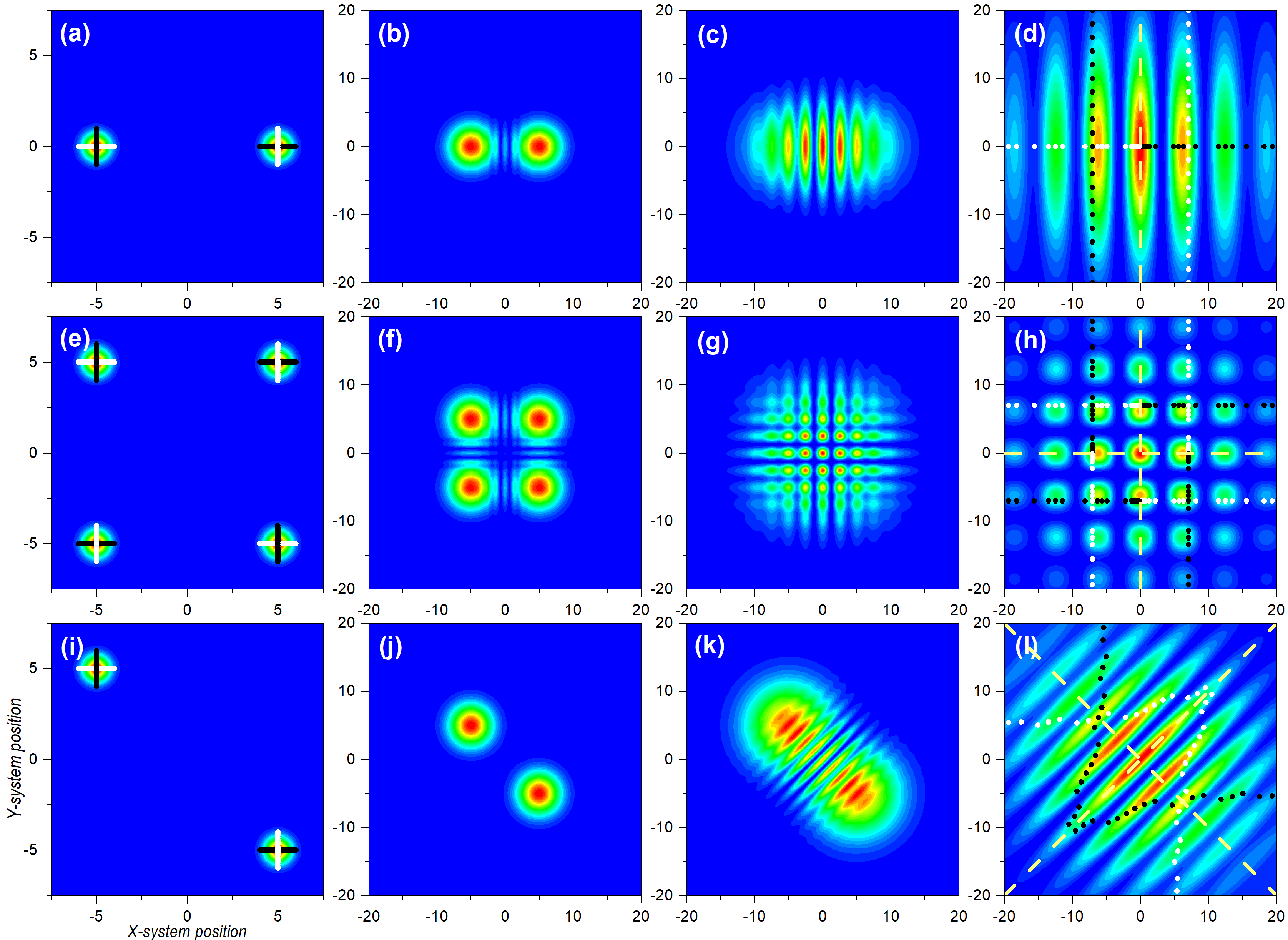}
 \caption{\label{Fig3}
  {Contour} 
 plots illustrating several stages of the evolution of the probability
  density for three bipartite systems.
  \textbf{Upper} \textbf{row}: Uncorrelated bipartite state described by a two-Gaussian
  superposition for $X$ and single Gaussian for $Y$.
  \textbf{Central} \textbf{row}: Uncorrelated bipartite state described by a two-Gaussian
  superposition for both $X$ and $Y$.
  \textbf{Lower} \textbf{row}: Entangled bipartite state described by a Bell-type state.
  From \textbf{left} to \textbf{right}: $t = 0$, $t = 2$, $t = 4$, and $t = 10$.
  {Color code: minimum density (zero) is denoted with blue and maximum with red in all cases.}
  In the first column panels, cross-shaped sets of markers superimposed to each Gaussian distribution
  denote the ensembles of initial conditions considered in the calculation of Bohmian trajectories.
  {Each horizontal/vertical ensemble contains a total of 21 evenly distributed initial
  conditions, chosen as follows: for the horizontal ensembles, within the interval $(x_c \pm \Delta x,y_{c'})$;
  for the vertical ensembles, within the interval $(x_c,y_{c'} + \Delta y)$.
  In all cases, $\Delta x = \Delta y = 1$, while we have $y_{c'} = 0$ for the single Gaussians describing $Y$
  in the upper panels, and $x_c, y_{c'} = x_A, x_B$, with $x_A = - x_B = 5$, for the superposition and the
  entangled states.}
  The markers in the fourth column panels indicate the final position of the corresponding Bohmian
  trajectories.\linebreak The numerical values used in the simulations were: $x_0 =0$ for the single Gaussian and $d = 10$
  for the other cases with two Gaussians, $\sigma_0 = 0.5$, $p_0 = 0$, $m=1$, and $\hbar = 1$.
  See text for further details.}
\end{figure}

\begin{figure}[!t]
\centering
\includegraphics[width=\textwidth]{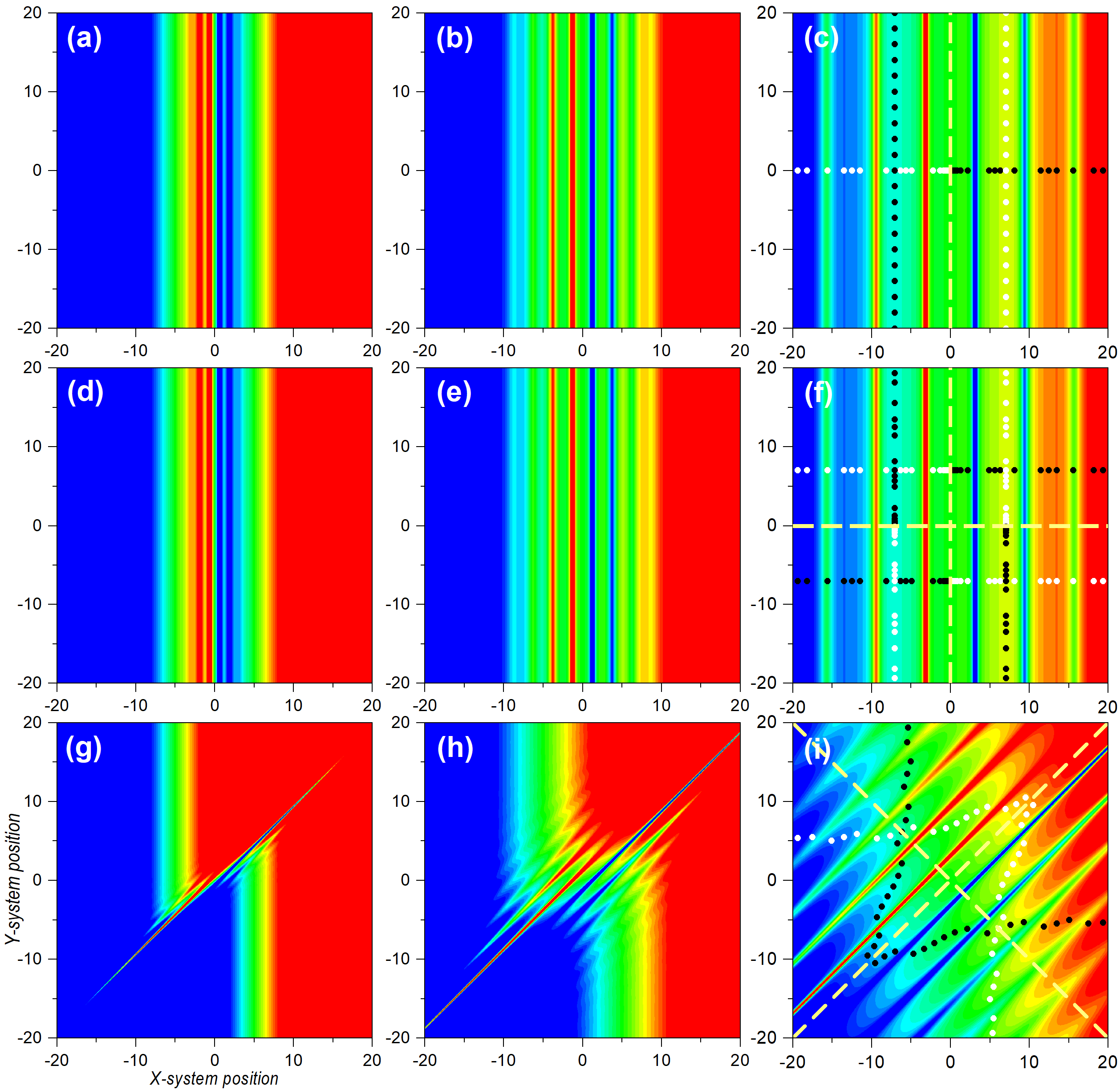}
 \caption{\label{Fig4}
  {Contour} 
 plots illustrating several stages of the evolution of the
  $x$-component of the transverse velocity field, $v^X(x,y|t)$, for the
  three bipartite systems of Figure~\ref{Fig3}.
  \textbf{Upper} \textbf{row}: Uncorrelated bipartite state, described by a two-Gaussian
  superposition for $X$ and a single Gaussian for $Y$.
  \textbf{Central} \textbf{row}: Uncorrelated bipartite state, described by a two-Gaussian
  superposition for both $X$ and $Y$.
  \textbf{Lower} \textbf{row}: Entangled bipartite statem described by a Bell-type state.
  From \textbf{left} to \textbf{right}: $t = 2$, $t = 4$, and $t = 10$.
  {For a better visualization, both the maximum (positive) and minimum (negative) values of the velocity fields
  represented in the nine snapshots have been truncated to 1.5 and $-1.5$, respectively.
  In the color code, these values are represented with red and blue, respectively, while greenish hues denote
  zero (and around) velocity values.}
  The markers in the third column panels indicate the corresponding final position of the sets of Bohmian
  trajectories referred to in the first column panels of Figure~\ref{Fig3}.
  The numerical values used in the simulations were: $x_0 =0$ for the single Gaussian and $d = 10$
  for the other cases with two Gaussians, $\sigma_0 = 0.5$, $p_0 = 0$, $m=1$, and $\hbar = 1$.
  \mbox{See text for further details.}}
\end{figure}

\begin{figure}[!t]
 \centering
 \includegraphics[width=\textwidth]{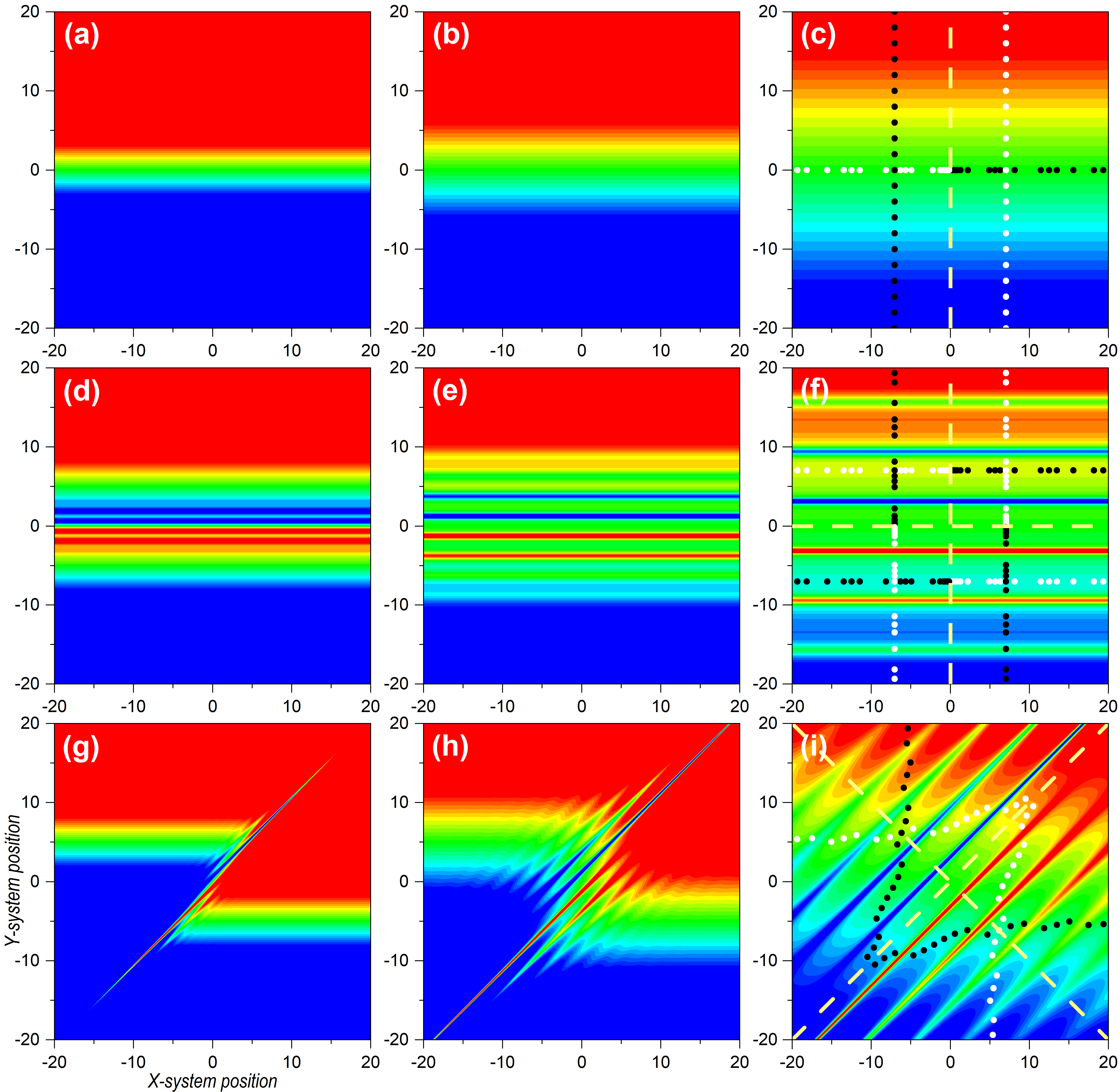}
 \caption{\label{Fig5}
  {Contour} 
 plots illustrating several stages of the evolution of the
  $y$-component of the transverse velocity field, $v^Y(x,y|t)$, for the
  three bipartite systems of Figure~\ref{Fig3}.
  \textbf{Upper} \textbf{row}: Uncorrelated bipartite state, described by a two-Gaussian
  superposition for $X$ and single Gaussian for $Y$.
  \textbf{Central} \textbf{row}: Uncorrelated bipartite state, described by a two-Gaussian
  superposition for both $X$ and $Y$.
  \textbf{Lower} \textbf{row}: Entangled bipartite state, described by a Bell-type state.
  From \textbf{left} to \textbf{right}: $t = 2$, $t = 4$, and $t = 10$.  
  {For a better visualization, both the maximum (positive) and minimum (negative) values of the velocity fields
	represented in the nine snapshots were truncated to 1.5 and $-1.5$, respectively.
	In the color code, these values are represented with red and blue, respectively, while greenish hues denote
	zero (and around) velocity values.}
  The markers in the third column panels indicate the corresponding final position of the sets of Bohmian
  trajectories referred to in the first column panels of Figure~\ref{Fig3}.
  The numerical values used in the simulations were: $x_0 =0$ for the single Gaussian and $d = 10$
  for other cases with two Gaussians, $\sigma_0 = 0.5$, $p_0 = 0$, $m=1$, and $\hbar = 1$.
  \mbox{See text for further details.}}
\end{figure}

Let us now consider the dynamics generated by bipartite separable and entangled
states, and, in particular, how the velocity field that rules the dynamics is influenced
by these states. The latter has important implications regarding the spatial
distribution of the quantum flux for each subsystem, and, hence, the corresponding
(reduced) probability densities and the observation of interference traits.
Unlike the analysis presented in Section~\ref{sec31}, here, the joint dynamics does not
facilitate an analogous space--time description, so we consider a
series of illustrative snapshots instead, for both the (joint) probability density and
the transverse velocity field.
Actually, in the latter case, given that this field depends on both subsystem
coordinates $x$ and $y$, we need to consider separate plots for the velocity fields
$v_x(x,y|t)$ and $v_y(x,y|t)$,
{specified, respectively, by Equations~(\ref{eq52}) and (\ref{eq53}).
}
Regarding the trajectories, we consider the trajectories projected onto the
corresponding subspaces (full higher-dimensional pictures of the trajectories within the
joint space are given in the Supplementary Material).
Specifically, in all cases, to investigate the dynamics generated by the corresponding
joint wave functions, we considered sets of trajectories with equidistant initial
conditions that mapped (initially) either the state of $X$ or the state of $Y$, and that
played the role of flow markers.
These initial conditions (21 for each wave packet) are uniformly distributed around
the center of the wave packets.
In the contour plots, these flow markers appear as dots with different colors along the
$x$ and $y$ directions for better identification in the long-time regime.
Thus, at $t=0$, they appear as a cross-shaped distribution (see first column panels in
Figure~\ref{Fig3}), while at $t=10$ they distribute in different ways (see last column panels
in Figures~\ref{Fig3}--\ref{Fig5}).

In the three cases analyzed, the $X$ subsystem was described by a two-Gaussian
coherent superposition, as given by Equation~(\ref{eq24}), as a simplified version of
Young's two-slit experiment.
In the case of the $Y$ subsystem, we firstly considered a localized state, such as the
Gaussian wave packet (\ref{eq14}).
Accordingly, the joint wave function read as
\begin{equation}
 \Psi(x,y|t) = \mathcal{N} \left[ \mathcal{G}_A(x,t) + \mathcal{G}_B(x,t) \right] \mathcal{G}_0(y,t) ,
 \label{eq58}
\end{equation}
The initial probability density is represented in Figure~\ref{Fig3}a, which effectively
shows that, while $Y$ was localized around $x_0$, $X$ was delocalized between $x_{A,0}$ and
$x_{B,0}$.
In Figure~\ref{Fig3}b--d we observe the evolution of the
joint probability density at $t=1$, $t=5$, and $t=10$, respectively.
Due to the factorizability of the state, we could easily identify the different stages
of the evolution for $X$ and $Y$ with the corresponding cases, shown in Figure~\ref{Fig1}a,d.
The spatial dispersion of the markers was also in compliance with the corresponding subsystems.
If we consider the markers distributed along the $y$-direction, their mutual separations
increased homogeneously, as this corresponded to the Gaussian state analyzed in Section~\ref{sec31}.
On the other hand, because of the emergence of interference traits, if we consider the markers
along the $x$-direction, their distribution was uneven, showing accumulations in the regions
covered by maxima and voids in the regions corresponding to interference minima,
separated by a distance $\Delta x = 2\pi$ at $t=10$, according to Equation~(\ref{eq35}).
These dynamics, though, were governed by the underlying velocity fields, as seen in the upper
row panels in Figures~\ref{Fig4} and \ref{Fig5}, for the $v^X(x,y|t)$ and $v^Y(x,y|t)$, respectively.\linebreak 
In particular, in Figure~\ref{Fig4}a--c, we observe the
snapshots for $v^X(x,y|t)$ at $t=1$, $t=5$, and $t=10$, respectively, while the analogous
cases for $v^Y(x,y|t)$ are represented in Figure~\ref{Fig5}a--c.
It is worth noting that, unlike the probability density, which was spatially modulated by the
Gaussian wave packets, decaying to zero at some point in either direction, the velocity fields
remained independent of the complementary coordinate.
Thus, at time $t$, $v^X(x,y|t)$ kept the same value for a given $x$-position, regardless of the value for $y$.
The same held for $v^Y(x,y|t)$ with respect to $x$.
This is particularly interesting in the case of $X$, where its associated velocity field showed,
for any value of $y$, the periodic alternation of nearly constant regions, in spatial intervals
with a value $\Delta x = 2\pi$ at $t=10$, again in compliance with Equation~(\ref{eq35}).

If we consider the case where both subsystems, $X$ and $Y$, are described by identical
delocalized superpositions,
\begin{equation}
 \Psi(x,y|t) = \mathcal{N}^2 \left[ \mathcal{G}_A(x,t) + \mathcal{G}_B(x,t) \right]
  \left[ \mathcal{G}_A(y,t) + \mathcal{G}_B(y,t) \right] ,
 \label{eq59}
\end{equation}
the situation turns a bit more complicated, because now the effects
of interference emerge along both directions, $x$ and $y$, giving rise to the appearance
of a highly organized chessboard-like pattern of local maxima modulated by a two-dimensional
Gaussian\linebreak  envelope, as seen in Figure~\ref{Fig3}h.
This can be inferred from the direct product between Equation~(\ref{eq32}) and its equivalent for
the $Y$ subsystem:
\begin{equation}
 \rho(x,y|t) = \rho(x,t) \rho(y,t) \approx 4 \left( \frac{2 m^2 \sigma_0^2}{\pi \hbar^2 t^2} \right)
   e^{-2m^2\sigma_0^2 (x^2 + y^2)/\hbar^2 t^2} \cos^2 (k_\infty x/2) \cos^2 (k_\infty y/2) .
 \label{eq60}
\end{equation}
A series of snapshots illustrating the trend, until the two-dimensional interference\linebreak  chessboard-like
pattern has reached full visibility, is shown in Figure~\ref{Fig3}e--h.
It can be noticed that, as in the previous case, sets of marker trajectories starting with the same
value of $x(0)$ or $y(0)$ do not mix (cross) with trajectories launched with other initial conditions,
in spite of the intricate pattern describing the asymptotic behavior of the probability density.
{The different colors assigned to the markers along the $x$ and $y$ directions for the
trajectories associated with the four different distributions make evident this behavior.} 
Nonetheless, the dynamical explanation can be found again in the underlying structure displayed
by the velocity field, shown in Figure~\ref{Fig4}e--h for $v^X(x,y|t)$, and in
Figure~\ref{Fig5}e--h for $v^Y(x,y|t)$.
As before, the factorizability of the joint state gives rise to two totally independent velocity
fields, although both have the same structure, which asymptotically displays the known alternation
of stable regions separated by sudden changes in the velocity at even distances of $\Delta x = 2\pi$,
for $t=10$.
According to this velocity pattern, the markers start acquiring an uneven spatial distribution as
they become more and more dispersed.

\begin{figure}[!t]
 \centering
\includegraphics[width=\textwidth]{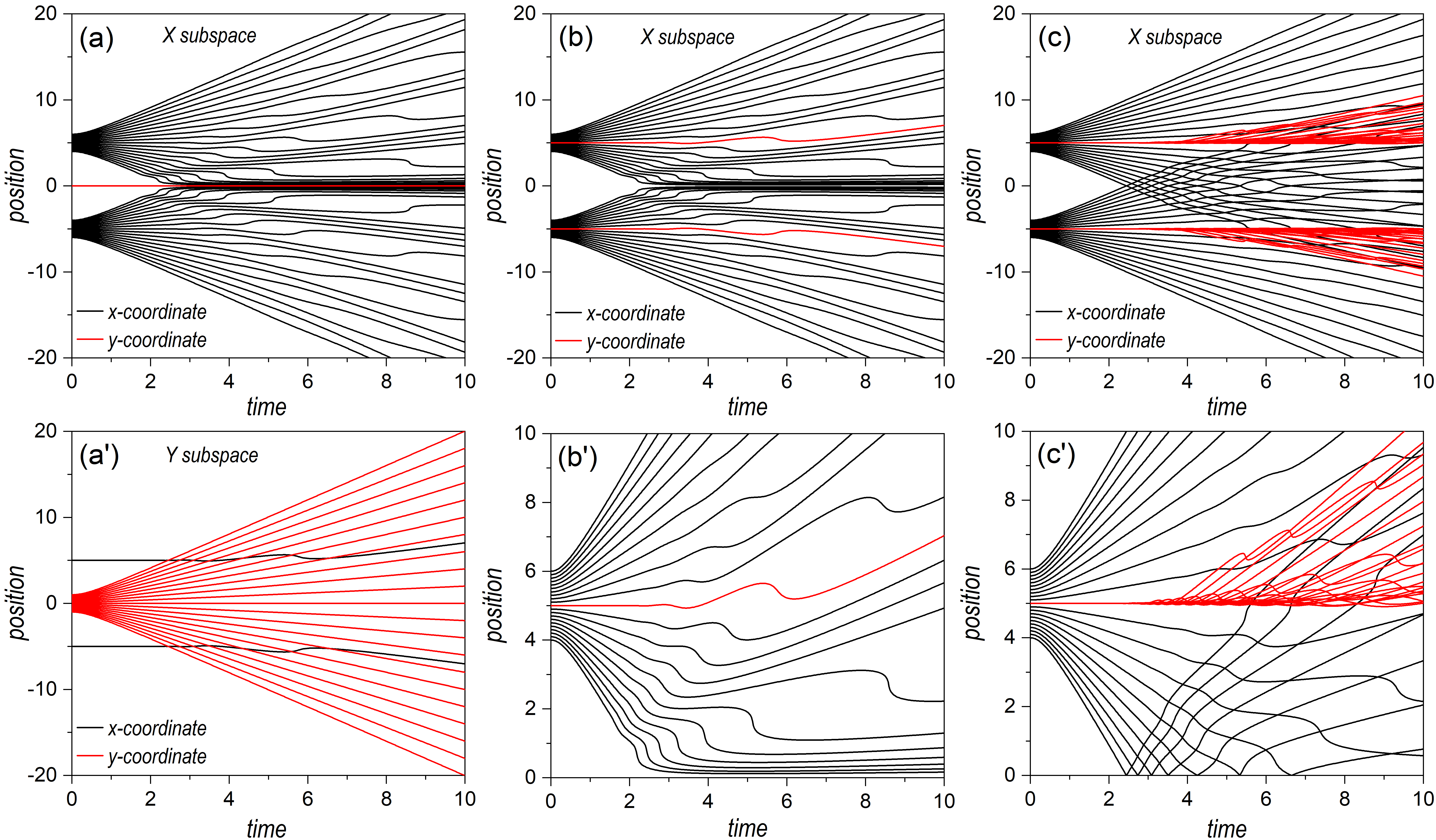}
 \caption{\label{Fig6} {Bohmian} 
 trajectories illustrating the dynamics associated with the three
  bipartite systems of Figure~\ref{Fig3}.
  \textbf{Left} \textbf{column}: Uncorrelated bipartite state, described by a two-Gaussian
  superposition for $X$ and single Gaussian for $Y$.
  The trajectories for each subsystem are plotted in the \textbf{top} and \textbf{bottom} \textbf{panels}, respectively.
  \textbf{Central} \textbf{column}: Uncorrelated bipartite state, described by a two-Gaussian
  superposition for both $X$ and $Y$.
  An enlargement of the upper half of the system is shown in the \textbf{bottom panel}.
  \textbf{Right} \textbf{column}: Entangled bipartite state described by a Bell-type state.
  An enlargement of the upper half of the system is shown in the \textbf{bottom} \textbf{panel}.
  As is shown in the corresponding first column panels in Figure~\ref{Fig3}, the initial conditions were chosen considering $21$ equidistant positions
  {either along the $x$ or the $y$ directions, covering the Gaussian wave packets in the manner
  specified, with more detail, in the caption for that figure.}
  In all plots, the $x$-component of the trajectories is represented with a solid black line, and the
  $y$-component with a solid red line.
  The numerical values used in the simulations were: $x_0 =0$ for the single Gaussian and $d = 10$
  for other cases with two Gaussians, $\sigma_0 = 0.5$, $p_0 = 0$, $m=1$, and $\hbar = 1$.
  See text for further details.}
\end{figure}

Next, let us take a look at the topology displayed by the trajectories followed
by the markers considered in the two cases analyzed so far.
A graphical representation on the dynamics for the three types of bipartite states here
considered is provided in Figure~\ref{Fig6}.
In the first column panels, we consider projections of the $x$ and $y$ trajectories
associated with the wave function (\ref{eq58}).
In Figure~\ref{Fig6}a, we have the $x$ and $y$-components of trajectories specified with
initial conditions of the type $[x(0),0]$ for the $X$ subsystem, with $x(0)$ evenly distributed
within the two regions covered by the two wave packets along the $x$-direction {(}see horizontal
markers in Figure~\ref{Fig3}a{)}.
As can be seen, while the $x$-component (solid black lines) was in agreement with the
behavior exhibited by the analogous one-dimensional system, described by Equation~(\ref{eq24})
and discussed in Section~\ref{sec31}, the $y$-component (red solid line) remained constant
[$y=y(0)=0$] at any time.
For any other value of $y(0)$, we would have obtained a similar behavior, with all trajectories
showing an identical behavior (for the same initial condition) along $x$, but with the whole
set moving along the vertical in compliance with the trajectory corresponding to a single
Gaussian (see Supplementary Materials).
In {Figure~\ref{Fig6}a$'$,}  
 however, the behavior is a bit different.
In this case, the initial conditions were of the type $[x_{A/B,0},y(0)]$, with $y(0)$ evenly
distributed within the region covered by a single wave packet along the $y$-direction {(}see
markers in Figure~\ref{Fig3}a along $y${).}\linebreak 
We observe that the $y$-component of the trajectories proceeded in agreement with the
behavior studied in Section~\ref{sec31} for a single Gaussian wave packet.
However, the $x$-component of the trajectories indicated wiggling behavior, since they were not
computed at $x(0)=0$, but at either $x(0)=x_{A,0}$ or $x(0)=x_{B,0}$, and, hence, there is going
to be a contribution to the motion of the $x$-component of the velocity field.
As a direct consequence of these behaviors, determined by inspecting the trajectory dynamics,
either in the $X$ subspace or in the $Y$ one, we note that there was a strong connection between
separability, a formal aspect, and diffusion, a physical behavior.
Specifically, if we consider a slice of the joint probability density at a constant value of
$x$ (or $y$), it looks the same at any time as that any other slice taken at another constant
value of $x$ (or $y$).
The associated (full) velocity field allowed such splittings at a dynamical level, while the
trajectories showed local (instantaneous) features of the diffusive process as time proceeded.
This consequence was also confirmed in the case wherein $X$ and $Y$ were both described by
delocalized wave functions, such as \mbox{in (\ref{eq59}}).
The corresponding trajectories ($x$ and $y$-components, as before) in the $X$-subspace are
plotted in Figure~\ref{Fig6}b.
In this case, due to the symmetry along both directions, the sets of trajectories with
initial conditions $[x_{A/B,0},y(0)]$ evolved following the same dynamics as their
counterparts with initial conditions $[x(0),y_{A/B,0}]$.
Consequently, we did not observe any crossing, as before, as can be more clearly seen in
the enlargement shown in {Figure~\ref{Fig6}b$'$.}
This resulted in the chessboard-type structure displayed by the joint probability density.

Let us now focus on the entangled state described by the wave function (\ref{eq42}).\linebreak 
As seen in Figure~\ref{Fig3}i, the initial probability density was only important around
$(d/2,-d/2)$ or around $(-d/2,d/2)$.
Physically, this meant that either $X$ went through slit $A$ and $Y$ through $B$, or
$X$ through $B$ and $Y$ through $B$, disregarding any other possibilities. In particular,
both subsystems passed through the same slit, which is what the diagonal terms,
$(d/2,d/2)$ or around $(-d/2,-d/2)$, describe.
The distance between centroids was, thus, larger than those between any two distributions in the
factorizable cases ($\sqrt{2} d$ for the entangled state, to be compared with $d$ for the factorizable state). Thus, the effects of the overlapping started becoming apparent at later
times {(no} effect is observed in Figure~\ref{Fig3}j, compared with the homologous separable cases shown in Figures~\ref{Fig3}b or \ref{Fig3}f{).}
Furthermore, because of the larger distance, the number of interference maxima increased and, hence, the distance
between minima decreased by a factor of $\sqrt{2}$.
These interferential traits, however, could only be seen along the diagonal joining the centroids, while no interference traits were apparent either along $x$ or $y$ (not, at
least, in a prominent manner).
This translated into interference suppression in the corresponding subspaces or
decoherence, discussed in Section~\ref{sec23}.

A priori, the above behavior could be considered a minor issue associated with
the non-separability of the joint state, since interference was still there, although along
the anti-diagonal direction joining the two centroids.
However, when we examined the corresponding velocity fields, $v^X(x,y|t)$ or $v^Y(x,y|t)$, displayed in Figures~\ref{Fig4}g--i and \ref{Fig5}g--i,
respectively, it was not possible to distinguish unique features associated
with one direction or the other since early stages of the time evolution.
Note that, when we look at these fields along the anti-diagonal direction, they are dramatically
interconnected at any time, without allowing us to clearly separate or distinguish two different contributions, remaining invariant along $x$ or $y$, as happened with the factorizable cases.
Of course, asymptotically we note that both velocity fields developed stable regions flanked by other regions with
sudden changes in the velocity, in compliance with the behavior observed in the probability density.
Yet, this is not enough to set an unambiguous separability criterion.
Accordingly, the markers disseminated in an uneven way everywhere {(}see Figure~\ref{Fig5}i{),}
which explicitly indicated the strong dependence that the dynamics of $X$ and $Y$ had on both
coordinates, $x$ and $y$.

If we now consider the trajectories associated with the markers of the entangled state, we
find a significant difference with respect to the separable states.
The trajectories associated with the $X$ subsystem are plotted in Figure~\ref{Fig6}c
(given the symmetry of the state, the trajectories for the $Y$ subsystem are identical).
As can readily be noticed, the renowned Bohmian non-crossing rule breaks down in the reduced subspaces as a consequence of the stretching of the velocity fields
along the direction perpendicular to the anti-diagonal.
The $x$-components of the trajectories launched with $y=0$ (solid black line) show clearly
how, beyond $t \approx 2$, crossing started, i.e., they passed through the same position at the
same time.
The $y$-components, on the other hand, instead of evolving along the same path as one of
the $x$-components, in analogy to the trajectories associated with the two-Gaussian
superposition {(}see {Figure~\ref{Fig6}b$'$),} also started diverting and crossing the same
point at the same time {(}see enlargement in {Figure~\ref{Fig6}c$'$).}
In this way, trajectories launched from the same subspace slice (i.e., the same
constant value for $y$, in the case of the $X$ subsystem) could now explore different regions (slices) of the joint space.
This is the reason why simply annihilating the interference term in a reduced model does
not ensure that trajectories cross the symmetry axis of the setup, as was shown
in \cite{sanz:EPJD:2007}, because the dynamics still contain contextual information on the
influence of both slits.
In order to find the correct answer, the reach of the information about the non-crossed
slit must be rather limited, or gradually suppressed, as shown in \cite{sanz:CPL:2009-2},
in compliance with the evolution here observed for the entangled trajectories.


\section{Concluding Remarks}
\label{sec4}

The above discussion poses an interesting question.
Typically, in a single-particle two-slit experiment there is no possibility to
distinguish which slit was crossed by the particle impinging on the detector.
However, Bohmian
trajectories provide us with a model, which tells us that there are immiscible streams
in the direction of the interference maxima that do not mix.
This is possible because there is an underlying velocity field that channels the swarms
of particles (probability flux) along given (quantized) directions, leaving voids between
every two neighboring ones.
Such velocity fields, as we have seen, are well-defined within the standard quantum
formalism (i.e., Schr\"odinger's picture), as is the link between the probability
density and the quantum flux, being two well-defined quantum quantities.
This is the reason why the experimental outcomes reported in \cite{kocsis:Science:2011},
and the trajectories inferred from them, are consistent.
In other words, a trajectory model is perfectly valid to describe quantum fluxes and is,
therefore, compatible with quantum mechanics (which does not mean that real particles
should be identified with the markers following the trajectories; this still remains
an empirically unproved matter).

Now, in the bipartite case, because of the wandering motion displayed by the tracers,
also mediated by the corresponding velocity field, the detection of particles associated
with the $X$ subsystem is ambiguous, since trajectories may come either from one
slit or the other, because the non-crossing rule is not valid anymore within the
reduced subspace.\linebreak This result explains why the experimental data reported in \cite{brida:PRA:2003} do not disprove the feasibility of a Bohmian picture, rather showing that it is compatible with standard quantum mechanics when the correct assumptions are considered about the (full) system analyzed. Note that, following the reasoning here, this is again a direct consequence arising from the underlying velocity field that relates the probability density with the quantum flux,
and not from additional postulates or interpretations. {In other words, the trajectory picture provided by Bohmian mechanics is consistent with standard quantum mechanics simply because all the main elements that it makes use of are contained within the latter; the trajectories only constitute an extra
refinement that allow us to determine, at a local level, the evolution of the probability
flux across the available phase space.}

{The discussion here has turned around bipartite Young-type interference experiments,
considering the particular case of Gaussian states because of analytical convenience.\linebreak 
However, we would like to stress that the method is applicable to a wide variety of problems
involving bipartite systems, because both the general expressions and the procedure are not
constrained to a specific representation.
In this way, it does not matter whether the two parties are spatially separated, as is
the case here, or, conversely, they are bound by an interaction potential.
The only difference with respect to the case here studied is the loss of analyzability, in
general, which requires resorting to numerical equations of motion.
}

\vspace{12pt}
\supplementary{The 
 following supporting information can be downloaded at: \url{https://www.mdpi.com/article/10.3390/e25071077/s1}
 (Alternatively, see Supplementary Materials section appended at the end of this document.)}


\vspace{6pt}
\noindent
{\small {\bf Funding:} This research was supported by the Spanish Research Agency (AEI) and the
European Regional Development Fund (ERDF) (Grant No.~PID2021-127781NB-I00P).}



\vspace{6pt}
\noindent
{\small {\bf Data Availability Statement:} The datasets generated during the current study are available
from the corresponding author upon reasonable request.}


\vspace{6pt}
\noindent
{\small {\bf Conflicts of Interest:} The author declares no conflict of interest.}






\reftitle{References}



\externalbibliography{no}



\newpage

\section*{Supplementary Materials}

\setcounter{figure}{0}
\renewcommand{\thefigure}{S\arabic{figure}}

The figures below show with more detail the dispersive effect led by the velocity fields $v^X(x,y|t)$ and
$v^Y(x,y|t)$ over swarms of 441 initial conditions distributed in an evenly-spaced $21\times 21$ square
array covering each Gaussian distribution and centered at the Gaussian centroid.
Moreover, the full three-dimensional trajectories corresponding to the cases analyzed in Fig.~6 (main
article) are shown in Fig.~\ref{Fig5SM} for a better understanding of the dynamics exhibited in the joint
($XY$) three-dimensional configuration space.

\begin{figure}[!h]
 \centering
 \includegraphics[width=0.5\textwidth]{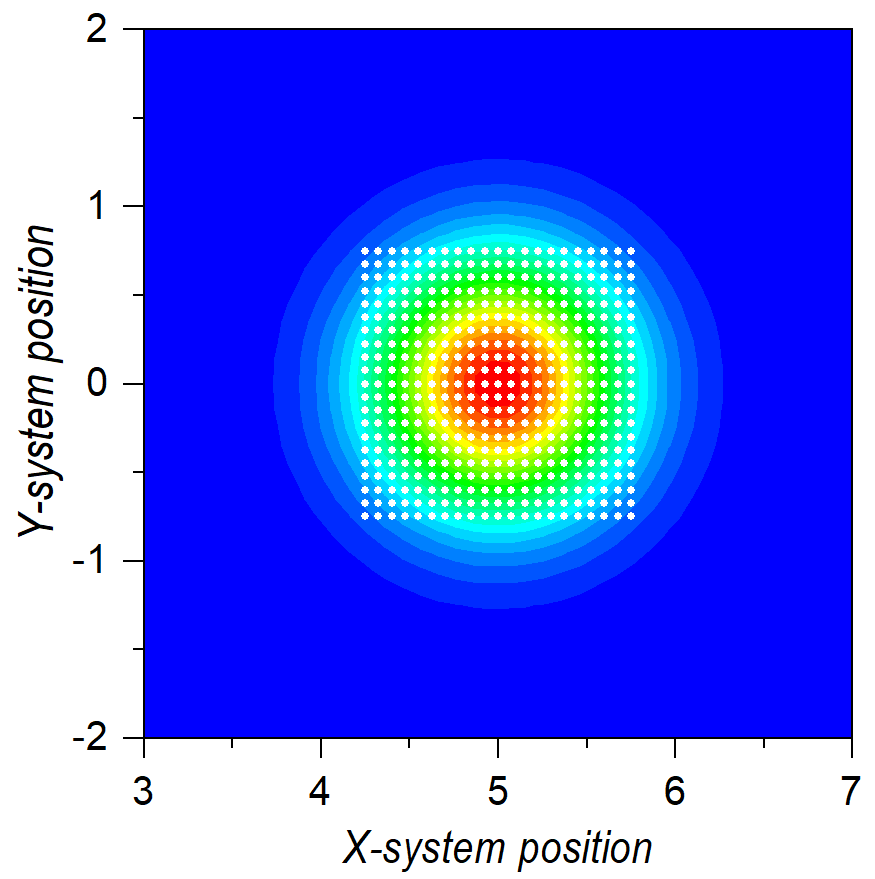}
 \caption{\label{Fig1Sm}
  Evenly-spaced $21\times 21$ squared array used to specify the initial conditions (solid white circles)
  of a set of 441 Bohmian trajectories.
  As an example, a single Gaussian distribution is shown, although the same array will be considered for
  all Gaussian distributions involved in each state analyzed.
  In this particular instance, the Gaussian distribution corresponds to the right-hand side wave packet of
  the $X$-subsystem superposition coupled to the $Y$-subsystem single Gaussian wave packet.
  The numerical values used here are: $\sigma_0 = 0.5$, $p_0 = 0$, $m=1$, and $\hbar = 1$.}
\end{figure}

\newpage

\begin{figure}[!t]
 \centering
 \includegraphics[width=\textwidth]{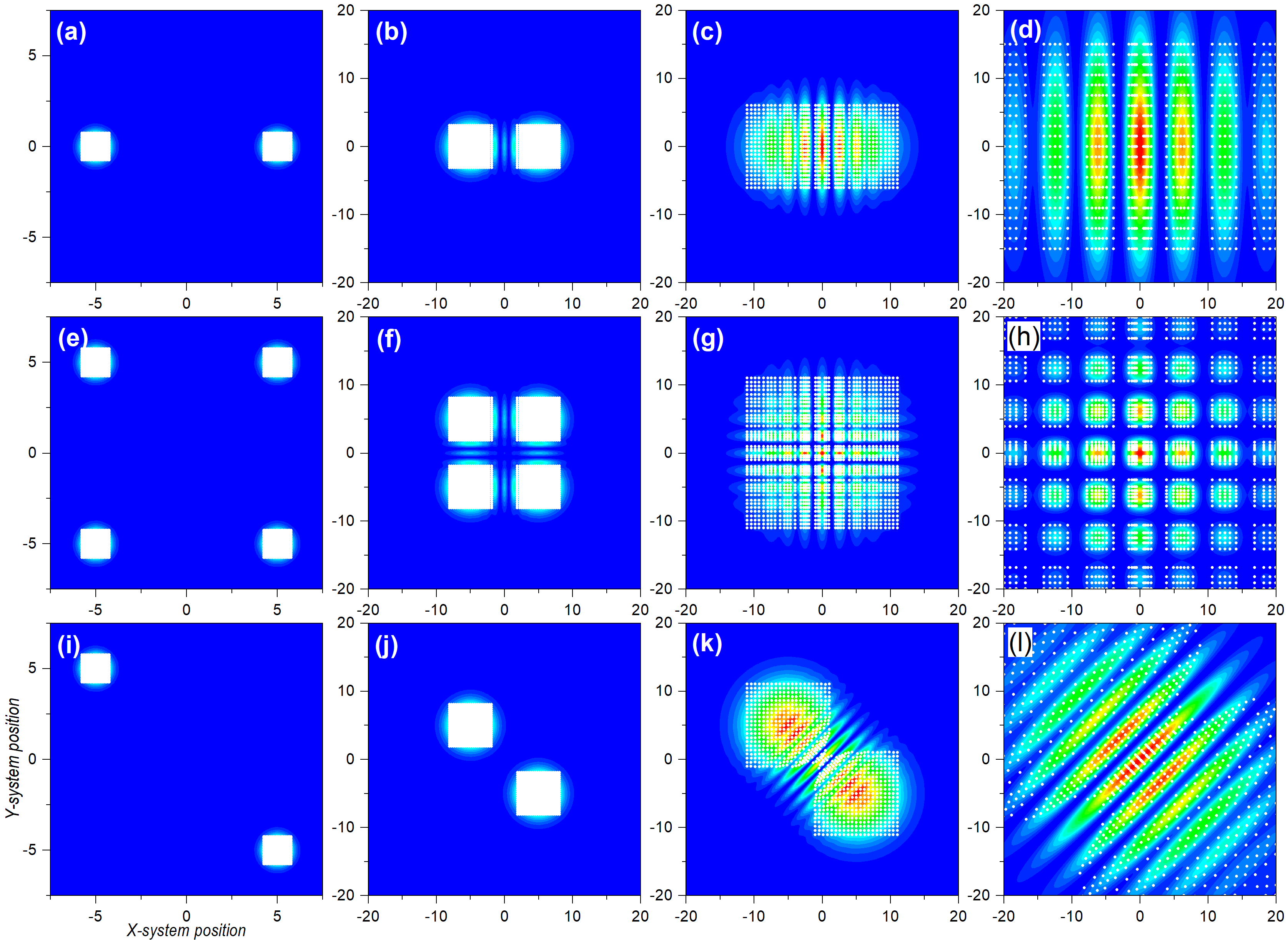}
 \caption{\label{Fig2SM}
  Contour plots illustrating several stages of the evolution of the probability
  density for three bipartite systems.
  Upper row: Uncorrelated bipartite state described by a two-Gaussian
  superposition for $X$ and single Gaussian for $Y$.
  Central row: Uncorrelated bipartite state described by a two-Gaussian
  superposition for both $X$ and $Y$.
  Lower row: Entangled bipartite state described by a Bell-type state.
  From left to right: $t = 0$, $t = 2$, $t = 4$, and $t = 10$.
  In the first column panels, the sets of 441 markers (solid white circles) distributed in squared arrays superimposed to each Gaussian distribution denote the swarms of initial conditions considered in the
  calculation of Bohmian trajectories.
  The markers in the next column panels show evidence of the difussion process that is taking place,
  which strongly depends on the different dyanmics generated by each joint state.
  The numerical values used in the simulations are: $x_0 =0$ for the single Gaussian and $d = 10$
  for other cases with two Gaussians, $\sigma_0 = 0.5$, $p_0 = 0$, $m=1$, and $\hbar = 1$.}
\end{figure}

\newpage

\begin{figure}[!t]
 \centering
 \includegraphics[width=\textwidth]{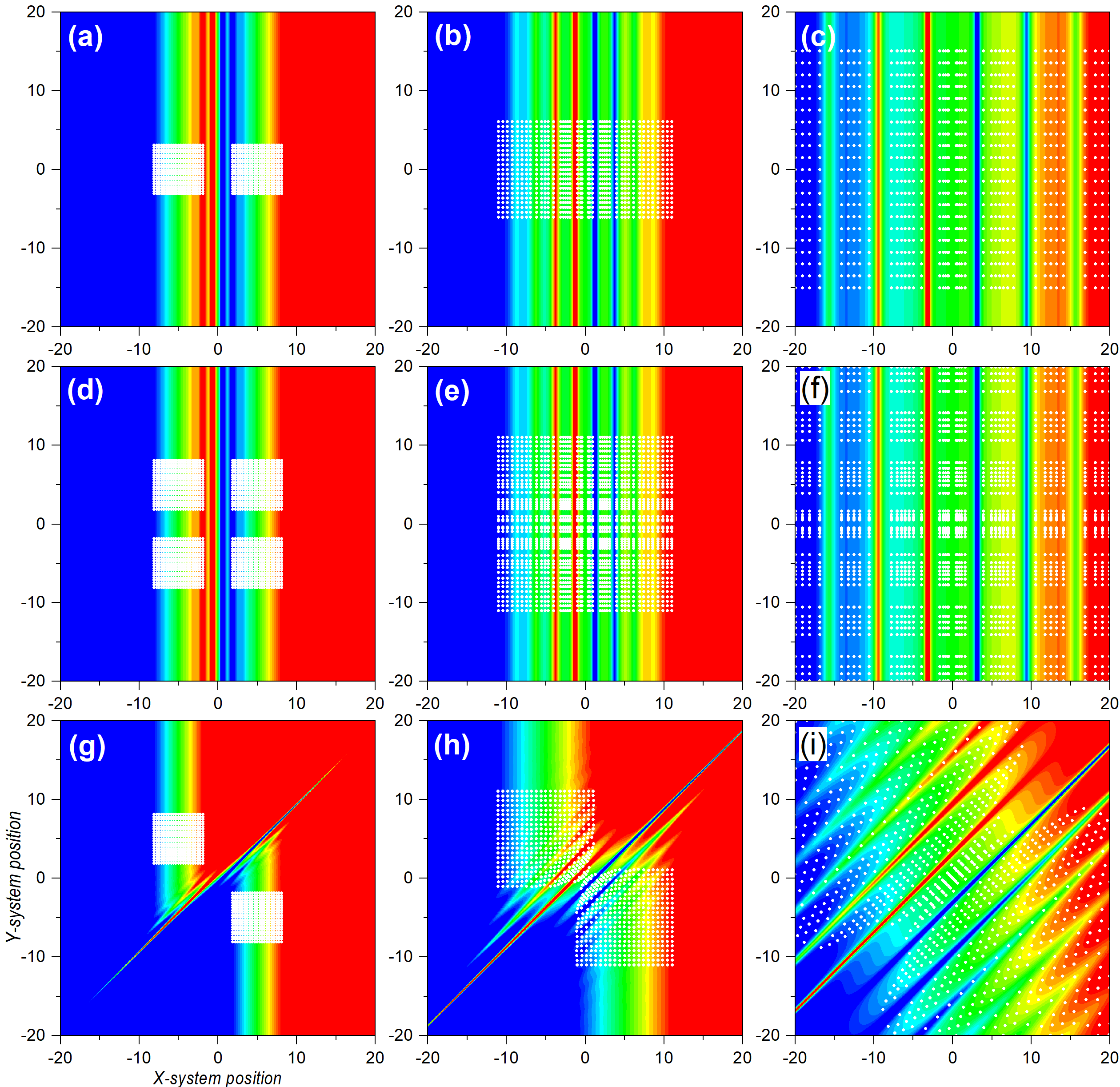}
 \caption{\label{Fig3SM}
  Contour plots illustrating several stages of the evolution of the
  $x$-component of the transverse velocity field, $v^X(x,y|t)$, for the
  three bipartite systems of Fig.~\ref{Fig2SM}.
  Upper row: Uncorrelated bipartite state described by a two-Gaussian
  superposition for $X$ and single Gaussian for $Y$.
  Central row: Uncorrelated bipartite state described by a two-Gaussian
  superposition for both $X$ and $Y$.
  Lower row: Entangled bipartite state described by a Bell-type state.
  From left to right: $t = 2$, $t = 4$, and $t = 10$.  
  The markers show how the velocity field directed along the $x$-direction is affecting the diffusion
  process of the trajectories in that direction.
  The numerical values used in the simulations are: $x_0 =0$ for the single Gaussian and $d = 10$
  for other cases with two Gaussians, $\sigma_0 = 0.5$, $p_0 = 0$, $m=1$, and $\hbar = 1$.}
\end{figure}

\newpage

\begin{figure}[!t]
 \centering
 \includegraphics[width=\textwidth]{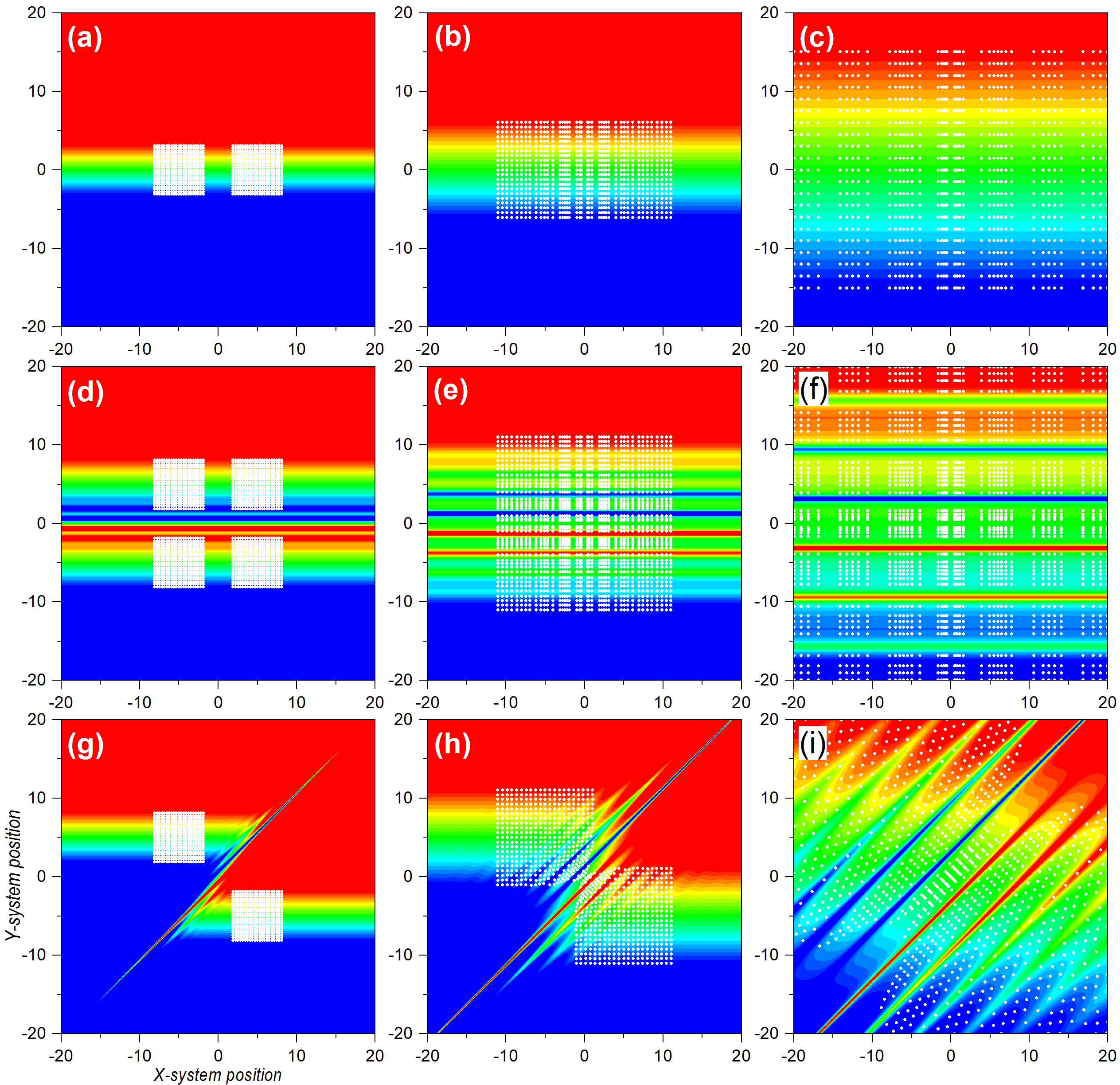}
 \caption{\label{Fig4SM}
  Contour plots illustrating several stages of the evolution of the
  $y$-component of the transverse velocity field, $v^Y(x,y|t)$, for the
  three bipartite systems of Fig.~\ref{Fig2SM}.
  Upper row: Uncorrelated bipartite state described by a two-Gaussian
  superposition for $X$ and single Gaussian for $Y$.
  Central row: Uncorrelated bipartite state described by a two-Gaussian
  superposition for both $X$ and $Y$.
  Lower row: Entangled bipartite state described by a Bell-type state.
  From left to right: $t = 2$, $t = 4$, and $t = 10$.  
  The markers show how the velocity field directed along the $y$-direction is affecting the diffusion
  process of the trajectories in that direction.
  The numerical values used in the simulations are: $x_0 =0$ for the single Gaussian and $d = 10$
  for other cases with two Gaussians, $\sigma_0 = 0.5$, $p_0 = 0$, $m=1$, and $\hbar = 1$.}
\end{figure}

\newpage

\begin{figure}[t]
 \centering
 \includegraphics[width=\textwidth]{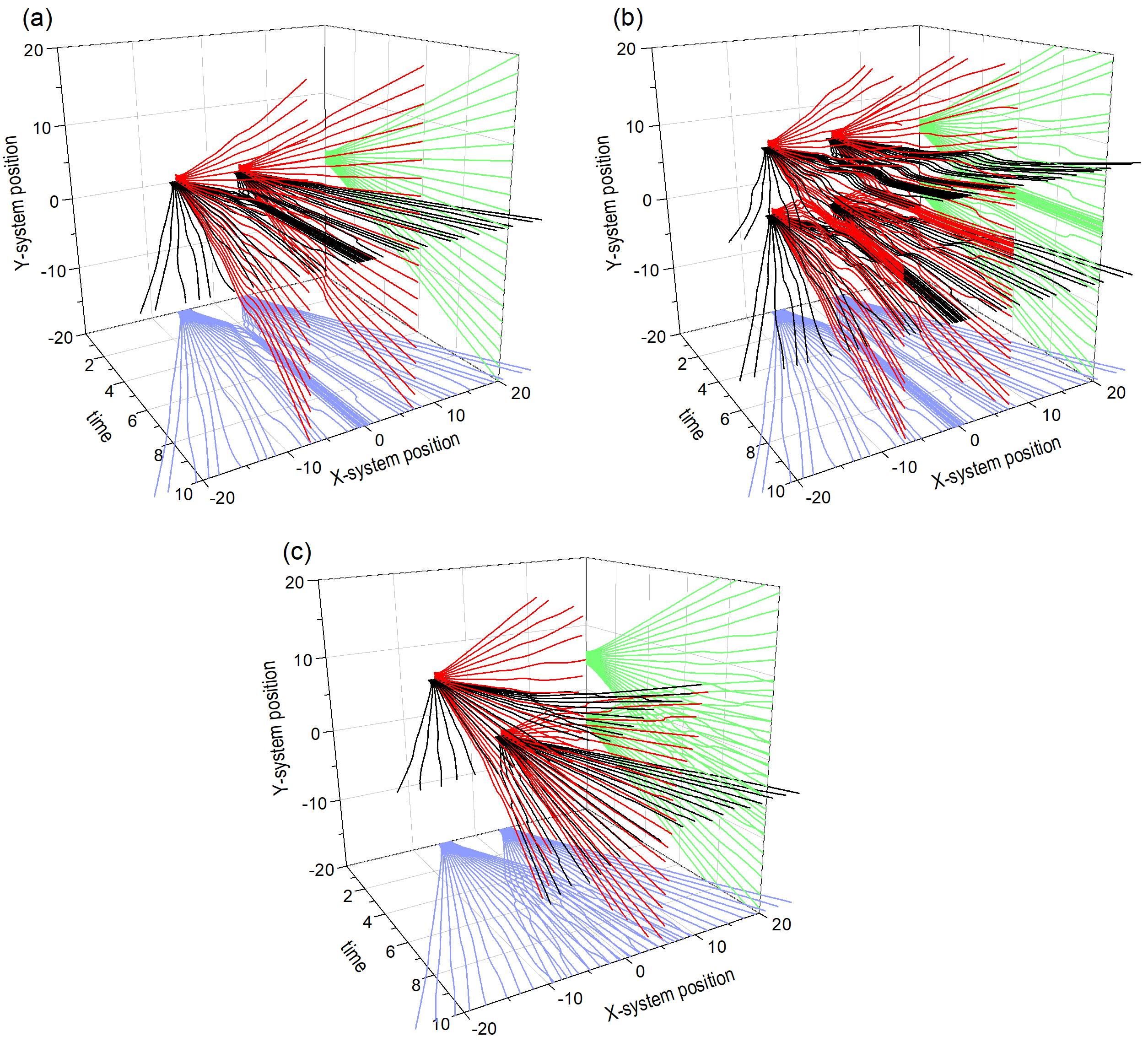}
 \caption{\label{Fig5SM}
  Three-dimensional representation of the Bohmian trajectories showed in Fig.~\ref{Fig6}, which
  illustrate the dynamics associated with the three bipartite systems of Fig.~\ref{Fig3}:
  (a) uncorrelated bipartite state described by a two-Gaussian
  superposition for $X$ and single Gaussian for $Y$;
  (b) the trajectories for each subsystem are plotted in the top and bottom panels, respectively;
  and (c) entangled bipartite state described by a Bell-type state.
  As it is shown in the corresponding first column panels in Fig.~\ref{Fig3}, the initial conditions
  have been chosen considering $21$ equidistant positions covering the Gaussian wave packet (along the $x$
  or the $y$ directions, depending on whether we are interested in the $X$ or the $Y$ subsystem,
  respectively).
  In all plots, the $x$-component of the trajectories is represented with solid black line, and the
  $y$-component with solid red line.
  The projections onto the bottom and side correspond to dynamics exhibited in the corresponding
  $X$ and $Y$ subspaces.
  The numerical values used in the simulations are: $x_0 =0$ for the single Gaussian and $d = 10$
  for other cases with two Gaussians, $\sigma_0 = 0.5$, $p_0 = 0$, $m=1$, and $\hbar = 1$.}
\end{figure}

\end{document}